\documentclass[a4paper,11pt]{article}
\pdfoutput=1 % if your are submitting a pdflatex (i.e. if you have
\usepackage{jcappub} % for details on the use of the package, please
\usepackage[T1]{fontenc} % if needed

\title{The PAU Survey: A Forward Modeling Approach for Narrow-band Imaging}

\author[a]{Luca Tortorelli,}
\author[a,b]{Lorenza Della Bruna,}
\author[a]{J\"org Herbel,}
\author[a]{Adam Amara,}
\author[a]{Alexandre Refregier,}
\author[c,d]{Alex Alarcon}
\author[c,d]{Francisco J. Castander,}
\author[e]{Juan De Vicente,}
\author[f]{Martin Eriksen,}
\author[g]{Enrique Fernandez,}
\author[h]{Juan Garc\'ia-Bellido,}
\author[c,d]{Enrique Gaztanaga,}
\author[g]{Ramon Miquel,}
\author[g]{Cristobal Padilla,}
\author[e]{Eusebio Sanchez,}
\author[c,d]{Santiago Serrano,}
\author[i]{Lee Stothert}
\author[f]{and Nadia Tonello}

\affiliation[a]{Institute for Particle Physics and Astrophysics, ETH Z\"urich, Wolfgang-Pauli-Str. 27, 8093 Z\"urich, Switzerland}
\affiliation[b]{Department of Astronomy, Oskar Klein Centre, Stockholm University, AlbaNova University Centre, SE-106 91 Stockholm, Sweden}
\affiliation[c]{Institute of Space Sciences (ICE, CSIC), Campus UAB, Carrer de Can Magrans, s/n, 08193 Barcelona, Spain}
\affiliation[d]{Institut d'Estudis Espacials de Catalunya (IEEC), E-08034 Barcelona, Spain}
\affiliation[e]{Centro de Investigaciones Energ\'eticas, Medioambientales y Tecnol\'ogicas (CIEMAT), Avenida Complutense 40, E-28040, Madrid, Spain}
\affiliation[f]{Port d'Informaci\'o Cient\'ifica, Campus UAB, c/ de l'Albareda, Edifici D, 08193 Barcelona}
\affiliation[g]{Institut de F\'isica d'Altes Energies (IFAE), Edifici Cn, Campus UAB, 08193, Barcelona, Spain}
\affiliation[h]{Instituto de Fisica Teorica, Universidad Autonoma de Madrid, Cantoblanco 28049 Madrid, Spain}
\affiliation[i]{Institute for Computational Cosmology, Department of Physics, Durham University, South Road, Durham DH1 3LE, UK}
\emailAdd{torluca@phys.ethz.ch}

\abstract{Weak gravitational lensing is a powerful probe of the dark sector, once measurement systematic errors can be controlled. In \cite{refregier14}, a calibration method based on forward modeling, called \textit{MCCL}, was proposed. This relies on fast image simulations (e.g., UFig) \cite{berge13} that capture the key features of galaxy populations and measurement effects. The \textit{MCCL} approach has been used in \cite{herbel17} to determine the redshift distribution of cosmological galaxy samples and, in the process, the authors derived a model for the galaxy population mainly based on broad-band photometry. Here, we test this model by forward modeling the 40 narrow-band photometry given by the novel PAU Survey (PAUS). For this purpose, we apply the same forced photometric pipeline on data and simulations using \textsc{Source Extractor} \cite{bertin96}. The image simulation scheme performance is assessed at the image and at the catalogues level. We find good agreement for the distribution of pixel values, the magnitudes, in the magnitude-size relation and the interband correlations. A principal component analysis is then performed, in order to derive a global comparison of the narrow-band photometry between the data and the simulations. We use a `mixing' matrix to quantify the agreement between the observed and simulated sets of Principal Components (PCs). We find good agreement, especially for the first three most significant PCs. We also compare the coefficients of the PCs decomposition. While there are slight differences for some coefficients, we find that the distributions are in good agreement. Together, our results show that the galaxy population model derived from broad-band photometry is in good overall agreement with the PAUS data. This offers good prospect for incorporating spectral information to the galaxy model by adjusting it to the PAUS narrow-band data using forward modeling.}

\begin{document}
\maketitle
\flushbottom

\section{Introduction}
\label{section:introduction}

The discovery of cosmic acceleration is among the most important results in modern cosmology (for a review, see~\cite{frieman08}). Evidence for acceleration includes data from supernovae (SNe) type Ia, clusters of galaxies, large-scale structure (LSS) and the cosmic microwave background (CMB). The evidence for a new era of accelerated expansion leads to several possibilities, the main two being: either the validity of GR breaks down on cosmological scales and a new, more complete theory of gravity is required, or about 70 \% of the energy density of the universe is present in form of a fluid with large negative pressure, known as dark energy, the behaviour of which is consistent with that of a cosmological constant $\Lambda$. Over the past several decades, experimental evidence has led to the establishment of the $\Lambda$CDM model. The $\Lambda$CDM universe (see~\cite{frieman08}) is composed of about 70 \% dark energy, while the remaining consists of 5\% (ordinary) baryonic matter and 25\% cold dark matter. The nature of dark matter and how to interpret dark energy are clearly two of the most important open questions in cosmology today.

A promising way to study the dark sector of the Universe is weak gravitational lensing (see~\cite{refregier03} and~\cite{hoekstra08} for a review). Weak gravitational lensing, or cosmic shear, is the weak distortion induced in the shape of a background galaxy through lensing by LSSs along the line of sight to the observer. As the name suggests, it is a small effect and thus subject to systematic errors (for a detailed description of systematic effects, see~\cite{bridle09} or a summary in~\cite{amara11}). To deal with them, \cite{refregier14} propose a forward modelling approach based on Monte Carlo Control Loops (\textit{MCCL}) and relies on fast image simulations. Its first full end-to-end implementation is presented in \cite{kacprzak18}, while, recently, \cite{fagioli18} develop a forward modeling tool to generate galaxy spectra, applying it to SDSS spectroscopic data. The \textit{MCCL} method relies on a series of tests to determine whether simulations have the same statistical properties as the data, and whether they are robust to a change of input model parameters and therefore not sensitive to systematic uncertainties. The Ultra Fast Image Generator (\textsc{UFig}) presented in \cite{berge13} was developed for this purpose. It relies on simple models of galaxy properties that need to be calibrated using existing data. The model was calibrated in~\cite{herbel17} using Subaru Suprime-Cam data~\cite{capak2007,Taniguchi2007} of the COSMOS~\cite{Scoville2007} field and used to determine the redshift distribution n(z) of cosmological samples.

Being calibrated using broad-band (BB) photometric data, it is important to test the model against new data that contains more spectral information. Therefore, the main goal of this paper is to test the performance of the standard configuration of the model with parameters taken from the literature. To do so, in this work we analyse both the Subaru data already used in~\cite{herbel17} and data from the Physics of the Accelerating Universe Survey (see section~\ref{section:data}). The PAU Survey is a novel imaging survey that provides photometric information through a set of 40 narrow band (NB) filters. This new dataset contains more information than a broad-band survey and thus it allows us to test the galaxy population model of \cite{herbel17} in a new regime. We will use Subaru data, both real and simulated, to detect sources taking advantage of the depth of broad-band. The output catalogues will be used to create detection images for the PAU Survey data and to perform forced photometry on them.

The paper is structured as follows. Section~\ref{section:data} presents our data and summarizes the principles of the image simulator \textsc{UFig}. Section~\ref{section:methodology} reviews the methodology. Catalogue and image level diagnostics are presented in section \ref{section:diagnostics}. A principal component analysis (PCA) on the full galaxy sample and on the Luminous Red galaxies sample is performed in section \ref{section:results}. We draw our conclusions and give an outlook in section \ref{section:conclusions}. Throughout this work, we use a standard $\Lambda$CDM cosmology with $\Omega_{\mathrm{m}}$ = 0.3, $\Omega_{\Lambda}$ = 0.7 and H$_0$ = 70 km s$^{-1}$ Mpc$^{-1}$.

\section{Data and Model}
\label{section:data}

Performing photometry with NB filters is not trivial. As the name suggests, the narrowness of the waveband probed makes the detection of sources a challenge. To overcome this problem, we use broad-band images to detect sources, from which we create detection images using the image simulator \textsc{UFig}. These are then used to perform the individual photometric measurements on the NB images, having already known positions and sizes of the objects. In our case, Subaru images of the COSMOS field are used as broad-band images. 

In this section, we describe the Subaru data itself and the PAU Survey data we analyze and simulate to test the model in \cite{herbel17}. The simulation is performed with UFig.

\subsection{PAUS and Subaru data}

The Physics of the Accelerating Universe survey (PAUS)~\cite{marti14,stothert18} is a photometric galaxy survey designed to map the LSS of the universe in three dimensions up to i$_{\mathrm{AB}}$ < 22.5-23.0 in the NB filters and i$_{\mathrm{AB}}$ < 24 in the BB filters with a quasi-spectroscopic redshift precision of $\sigma(\mathrm{z})/(1+\mathrm{z}) \sim 0.0035$. The target fields of PAUS contains, among others, the COSMOS field, the CFHTLS W1, W2, and W3 fields \cite{cfhtls} and part of the GAMA fields \cite{GAMA}.

The survey is conducted using the PAU camera (PAUCam) \cite{padilla16,padilla18}, a large field of view camera installed at the William Herschel Telescope (WHT) in the Observatorio del Roque de los Muchachos in La Palma (Canary Islands, Spain). The camera is equipped with 18 CCDs that cover about 1 squared degree in the sky (the 8 central science CCDs cover about 1300 squared arcminutes, excluding the gaps between CCDs). The pixel scale varies across the focal plane, with an average of $0.265$ arcsec/pixel. PAUCam uses a set of 40 NB filters spanning the range 450 - 850 nm, each with a width of 12.5 nm at FWHM and overlapping one with each other\footnote{https://www.pausurvey.org/paucam/filters/}, together with a standard set of BB filters (u, g, r, i, z, Y). The NB filters are arranged on 5 filter trays, with 8 filters per tray, located on the central CCDs, while the 6 BB filters are placed on 6 trays, which cover all the CCDs\footnote{https://www.pausurvey.org/paucam/detectors/}. At each position on the sky, all the filter-trays are interchanged. Given a filter-tray, each of the 8 CCDs has a different NB filter. Therefore, to image a particular region of the sky in all the 40 NB filters, different observation patterns can be used. Of the various possible ones, a spiral pattern was chosen to cover the PAUS fields \cite{padilla18}.

In this work, we use PAUS data covering the COSMOS field. RA and DEC of these fields are reported in Appendix~\ref{appendix:pauscoordinates}.

As for the BB data, we use publicly available\footnote{http://cosmos.astro.caltech.edu} images of the COSMOS field taken with the Suprime-Cam on the 8-m Subaru telescope. We use images in the r+ band having their original seeing (no PSF homogeneization applied). The pixel scale is 0.15 arcsec/pixel and the total area covered on the sky in the COSMOS field is 1.86 deg$^2$.

\subsection{\textsc{UFig} and modifications for PAUS}
\label{section:UFig}

\begin{figure}[t!]
\centering
\includegraphics[width=4.5cm]{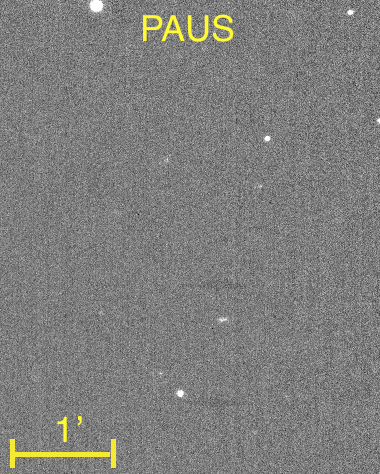}
\includegraphics[width=4.5cm]{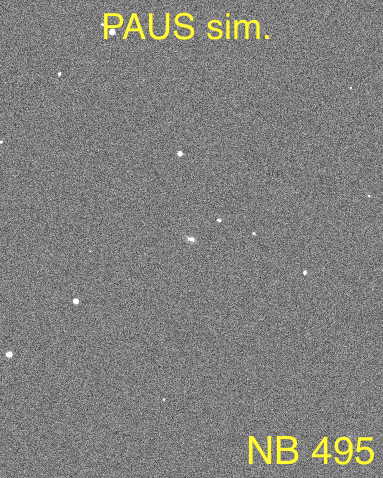} \\
\includegraphics[width=4.5cm]{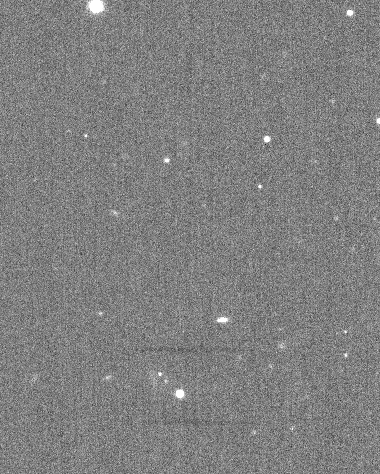}
\includegraphics[width=4.5cm]{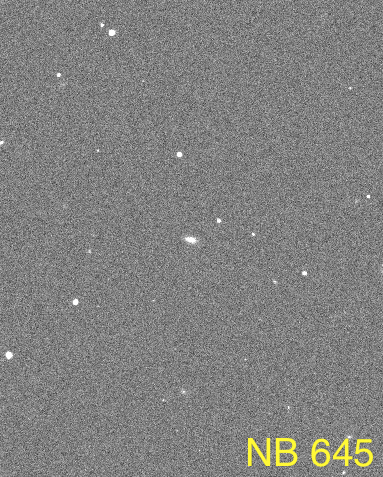} \\
\includegraphics[width=4.5cm]{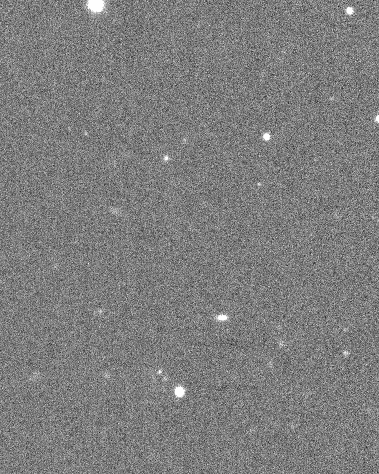}
\includegraphics[width=4.5cm]{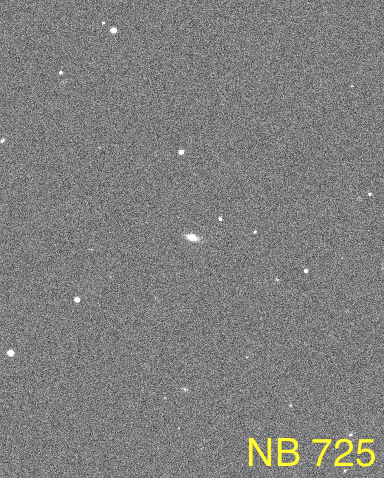}
\caption{PAUS observed and simulated images in the field RA: 09h 58m, DEC: 02$^{\circ}$ 05$^{'}$. The left panels show the observed PAUS image in the NB495, NB645 and NB725 filters (from the top to the bottom), while the right panels show the simulated PAUS image in the same NB filters. Each simulated image contains sources which are randomly distributed on the sky.}
\label{fig:tortorelli_fig1}
\end{figure}

This section summarizes the basic working principles of the image simulator, its underlying model and the main modifications we implement to simulate PAUS images. An extensive description of the simulator, of the model and its calibration are given in~\cite{berge13,herbel17}.

\textsc{UFig}~\cite{berge13} is a fast code able to generate simulated astronomical images in any arbitrary filter band, including the effect of redshifts and colors of galaxies. \textsc{UFig} first generates catalogues of sources, randomly drawing galaxies and stars from distributions of physical parameters of the galaxy population, and then renders their pixelated light profiles. To fully simulate an actual image, it also includes observational and instrumental effects, such as noise, PSF and saturation. 

\textsc{UFig} was primarly developed for forward modelling purposes. To be able to run the \textit{MCCL} (see section \ref{section:introduction}), where thousands of simulated images are needed, speed is an essential factor. Indeed, the runtime of \textsc{UFig} is comparable with the timescale of analysis softwares such as \textsc{Source Extractor}~\cite{bertin96} (hereafter, \textsc{SE}), e.g., less than one minute for 0.25 squared degrees. The simulator makes use of simplifying models for speed purposes. By running the \textit{MCCL}, it can be investigated whether more complex models are required.

By drawing from a given luminosity function $\Phi(M,z)$, \textsc{UFig} assigns to every galaxy a redshift and an absolute magnitude. The size and the total number of galaxies in an image are also modelled. Each galaxy is also assigned a spectrum, which is a linear combination of five basis spectra taken from \textsc{kcorrect} templates \cite{blanton07}. This particular way of assigning spectra and the coefficients of the combination are empirically motivated by SDSS data \cite{beare15}. These steps are repeated separately for two different luminosity functions for the populations of (young) blue galaxies and (old) red galaxies. Sources are then randomly distributed on the sky and they do not have the same position of PAUS real sources. Angular clustering is not present in the current implementation of UFig. The specific functioning point, i.e. the specific set of the model parameters we use in our work is described in Appendix \ref{appendix:functioningpoint}. We test the performance of this functioning point based on broad-band data against our new datasets.

UFig is a flexible code that can simulate different photometric dataset. We modify the original implementation of \textsc{UFig} in order to be able to simulate PAUS images. The modifications relate to how the image is rendered and to the catalogue of sources from which \textsc{UFig} renders the objects.

We modify the background generation and the astrometry. We simulate the background noise in every pixel of the PAUS image by randomly drawing from a Gaussian. The values of the noise mean and standard deviation are taken from the PAU database\footnote{http://www.pic.es} (PAUdb) \cite{tonello18} and they differ from exposure to exposure and from CCD to CCD. According to the image that we are simulating, we use the CCD number and the exposure number to find the correspondent values and then apply them to simulations.

The original implementation of \textsc{UFig} creates simulated images using the astrometric library in the Python module \textsc{Astropy}~\cite{astropy}. This however is not able to correctly handle the projection distortion parameters that are present in the header of PAUS images. To account for those, we switched to the Ast astrometric library\footnote{http://www.starlink.ac.uk/docs/sun211.htx/sun211.html\#xref\_}, which is the one used by the visualization software DS9\footnote{http://ds9.si.edu/site/Home.html} and GAIA skycat\footnote{http://star-www.dur.ac.uk/~pdraper/gaia/gaia.html}. The correct handling of the astrometry allows us to create images which are rotated w.r.t. the celestial North and distorted by the same amount as the real ones. Furthermore, it allows us to render galaxies in the correct position on the pixel grid. Indeed, ignoring the high-order polynomials would result in an object centroid shift of few arcseconds, which would have a significant impact on the forced photometry measurement. 

We use \textsc{UFig} to create the detection images for PAUS data and therefore we modify it to render objects which have photometric parameters (e.g., positions, rescaled size, template coefficients) taken from the Subaru data (see section \ref{section:methodology} for more details) and not randomly drawn from the luminosity function. The same holds for the detection images creation for PAUS simulated images. We read source parameters from Subaru simulated images and we render them on the PAUS simulated images in order to have the same source in the same position and with the same properties for both Subaru and PAUS. Also in the case of Subaru images, sources are randomly distributed on the sky.

Lastly, we also modify \textsc{UFig} to render objects on the PAUS images with the correct position angle with respect to the sky coordinates. The position angle of objects is determined from Subaru data as the rotation angle of the photometric semi-major axis with respect to the x-axis. Subaru images are aligned such that the y-axis corresponds to the celestial North direction, while for PAUS images it corresponds to the celestial East direction. Therefore, in order to match the position angles, we apply a rotation matrix to the Subaru estimate of them. The matrix first rotates the position angle according to the rotation angle $\cos{\theta} = \frac{\mathrm{CD1\_1}}{\mathrm{CDELT1}}$ and then flips it over the y-axis. The combination results in the matrix
\begin{equation}
\left[
\begin{matrix}
    -\cos{\theta}     & \sin{\theta} \\
    \sin{\theta}       & \cos{\theta} 
\end{matrix}
\right]
\end{equation}

In figure ~\ref{fig:tortorelli_fig1}, we show an example of PAUS simulated images in the field RA: 09h 59m, DEC: 02$^{\circ}$ 21$^{'}$. The left panels show the observed PAUS images in the NB495, NB645 and NB725 filters, while the right panels show the simulated PAUS image in the same NB filters, but with randomly distributed sources.

We simulate the 143 Subaru tiles covering the COSMOS field, using the same instrumental and astrometric parameters as the real ones. Then, by using the output properties of each rendered source (rescaled to take into account the different pixel scale), we simulate the 40 NB images for each PAUS field used. 

\section{Method}
\label{section:methodology}

\begin{figure}[t!]
\centering
\includegraphics[width=17cm]{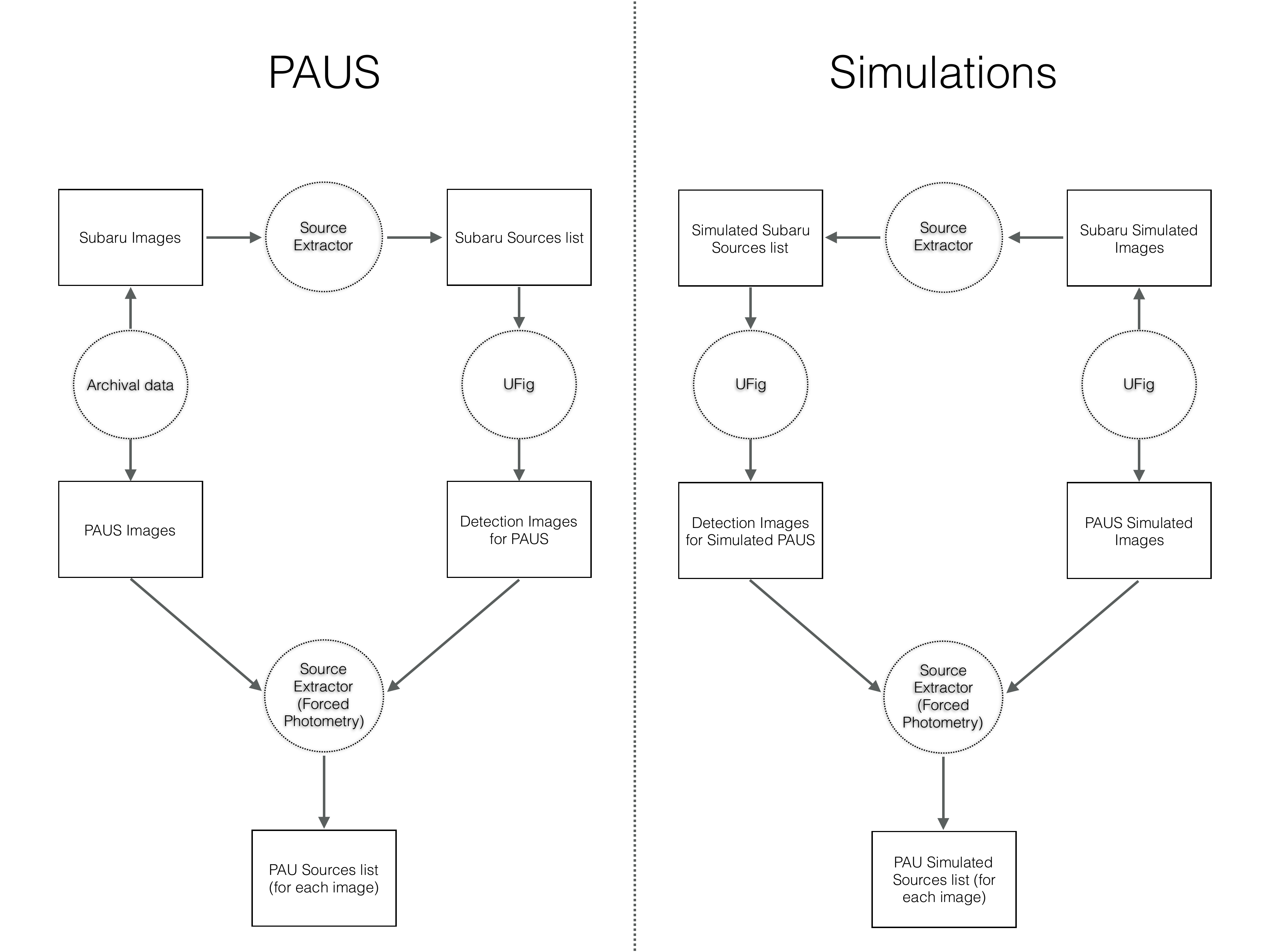}
\caption{Flowchart describing the fast analysis pipeline. On the left, we show the data analysis steps performed on real PAUS image, while on the right we show the same analysis steps, but performed on the simulations. Note that the simulations have different input parameters than the real images, but they are modeled to statistically match PAUS data.}
\label{fig:tortorelli_fig2}
\end{figure}

In this section, we describe the method (hereafter, pipeline) that we use to analyze and apply the forward modeling approach to PAUS images and test the model described in \cite{herbel17}. We apply the pipeline to the PAUS images belonging to the COSMOS field. We filter out multiple exposures and we analyze, in total, 2400 NB PAUS images. The scheme of the pipeline is illustrated in figure~\ref{fig:tortorelli_fig2}.

The data analysis can be divided into three main blocks: the detection catalogue generation from deep broad-band data, the forced photometry and the final catalogue creation. A crucial feature is the fact that the same analysis steps are applied to both data and simulations. Indeed, the forward modeling approach relies on producing realistic simulations that are analysed in the same way as the real data. Therefore, what we need is a fast pipeline that can be applied consistently to data and simulations.

\subsection{Detection Catalogue Generation}
\label{subsection:detectionbroad}

\begin{figure}[t!]
\centering
\includegraphics[width=13cm]{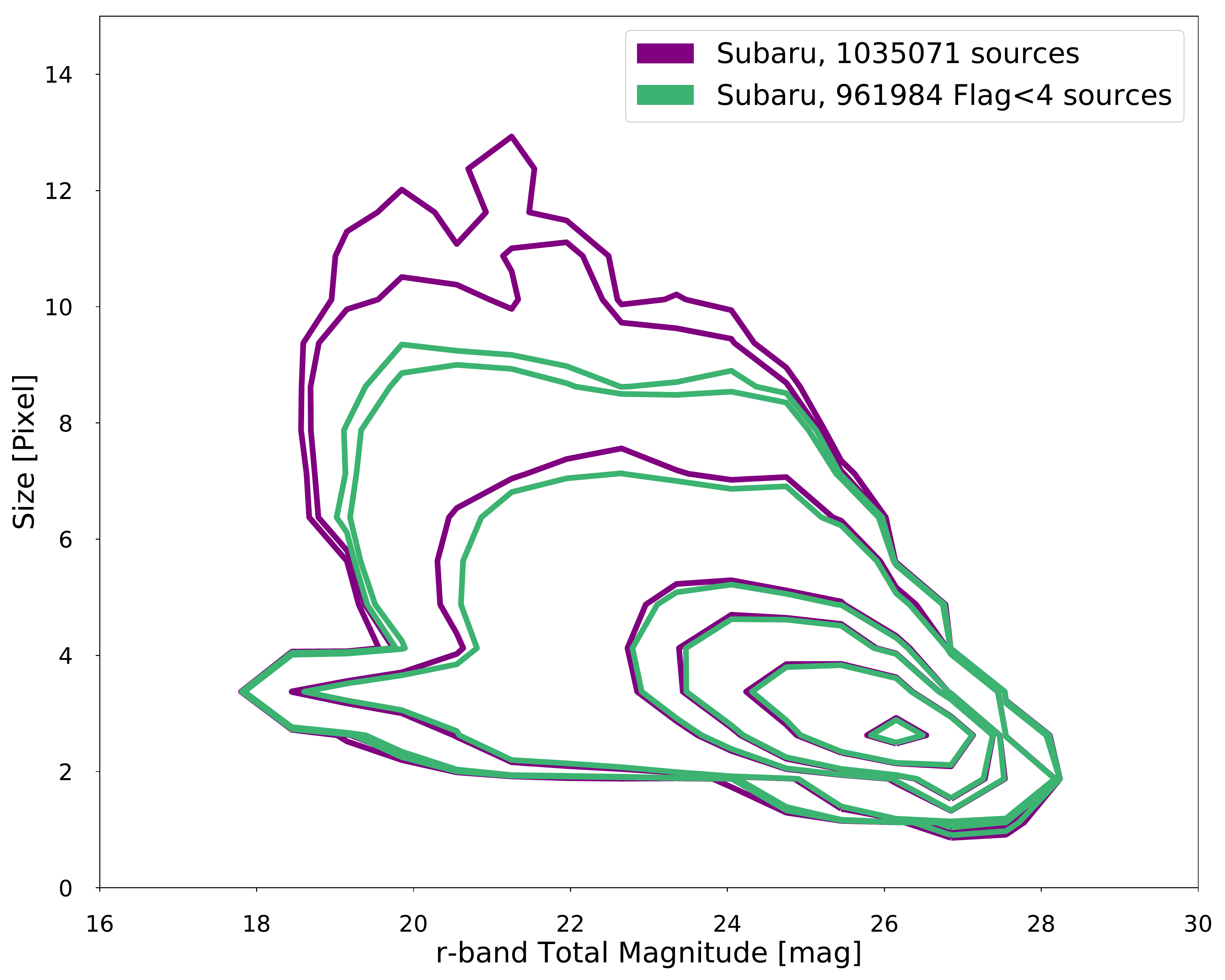}
\caption{Total magnitude in the r-band versus size in pixel. The plot shows the distribution of the total number of sources analysed in the 143 Subaru images. Purple contours represent all the objects extractred with \textsc{SE}. Green contours represent sources having \textsc{SE} `FLAGS' parameter less than 4. The upper right legend reports the number of sources for the two sets.}
\label{fig:tortorelli_fig3}
\end{figure}

As already highlighted in section~\ref{section:data}, the narrowness of the NB filters makes the detection of sources difficult. PAUS NB images have a limiting magnitude of i$_{\mathrm{AB}}$ $\sim$ 22.5, therefore they are shallow compared to BB images of deeper photometric surveys. Furthermore, integrating over a small wavelength range makes the detection of sources hard in certain NB filters, e.g., where a galaxy absorption line may fall. One way out is to use the broad-band Subaru data to detect sources and then use the resulting detection catalogue to create detection images with \textsc{UFig} and to perform forced photometry on PAUS data. We perform forced photometry with \textsc{SE}. We therefore need to provide a detection and a measurement image, since \textsc{SE} does not accept a detection catalogue. The real and the detection images must have the same size in order for the forced photometry to work.

First, we run \textsc{SE} on both Subaru real and simulated images. The used \textsc{SE} configuration is the same for real and simulated images and it is described in Appendix~\ref{appendix:sexconfig}. To avoid spurious detections and contaminations, we perform a cut that only select objects with a \textsc{SE} `FLAGS' parameter less than 4. Furthermore, we also cut sources which lie less than 40 pixels from bright saturated stars. For a typical observed Subaru tile, this cut corresponds to a decrease of about 10\% in the number of sources. We show the total magnitude-size plot for the Subaru images analysed in this work in figure~\ref{fig:tortorelli_fig3}. The cut exclude very bright stars (bright end of the fixed size locus), which are mostly saturated, and objects with very high values of size (half-light radius) in pixels. A visual inspection of these large size objects having \textsc{SE} `FLAGS' parameter greater than 4 shows that they are due to bad pixels caused by saturated stars.

Since UFig renders galaxy light distributions according to the S\'ersic profile, to create the detection images, besides positions and sizes, we estimate also the S\'ersic index from Subaru images. We obtain this through \textsc{SE} fits to each object. To do this, we need to provide \textsc{SE} with a PSF model for each Subaru image. Therefore, we select stars in the image using the GAIA DR1~\cite{gaia,gaiadr1}. We select only bright stars with magnitude in the g-band 19 < m < 20 and we cut 30$\times$30 pixel stamps around each star. We fit a 2D circular Moffat profile to every star to determine the seeing FWHM. The final seeing estimate is the mean of all the FWHM values. To create the model of the PSF used in \textsc{SE}, we combine the image stamps, after recentering, by averaging them.

\subsection{Forced Photometry in PAUS}

\begin{figure}[t!]
\centering
\includegraphics[width=4.7cm]{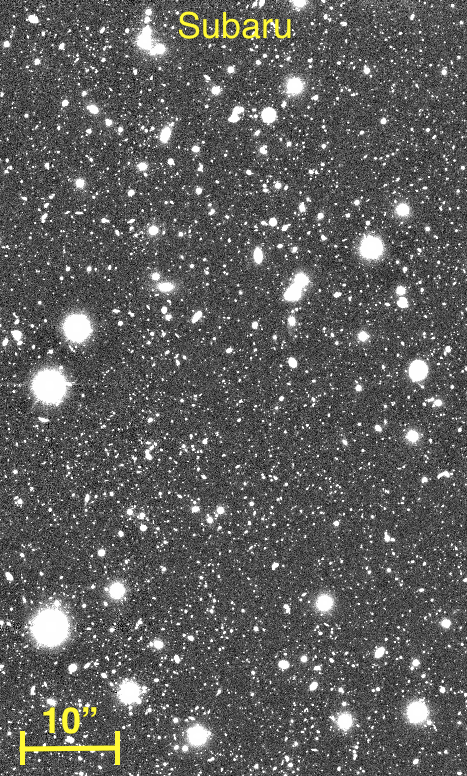}
\includegraphics[width=5.1cm]{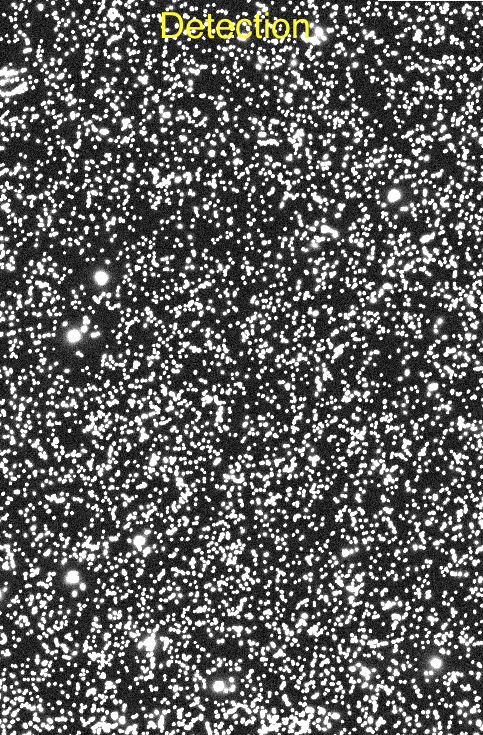}
\includegraphics[width=5.15cm]{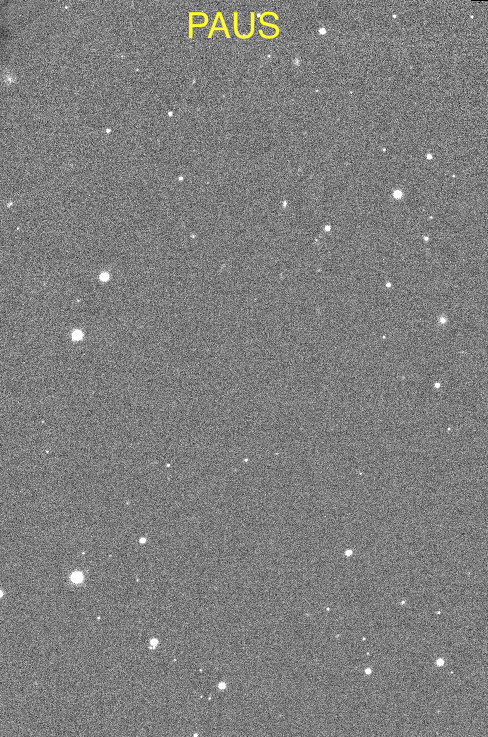}
\caption{The left panel shows a crop of an r-band Subaru image in the COSMOS field. The central panel shows the detection image, created using sizes and positions from the overlapping Subaru image, for the NB725 filter PAUS tile in the field RA: 09h 58m, DEC: 02$^{\circ}$ 05$^{'}$. The right panel shows instead the real NB725 filter PAUS image. The size of the Subaru crop is equal to that of the overlapping PAUS tile and detection image. The lack of sources in some regions of the detection image is due to the presence of bright stars which have \textsc{SE} `FLAGS' greater than 3. The larger number of sources visible in the central panel with respect to the Subaru one (left panel) is due to the fact that every detected source has a high S/N in the detection image to facilitate the photometry.}
\label{fig:tortorelli_fig4}
\end{figure}

Forced photometry is the preferable approach to measure fluxes and hence to compute galaxy properties from NB images. To perform it, we run \textsc{SE} in the dual image mode. Since \textsc{SE} uses the detection image only to find positions and apertures of objects, the detection images created with \textsc{UFig} are highly simplified. Sources are rendered on the image according to a S\'ersic profile, which has S\'ersic index, half-light radius, centroid position, axis ratio and position angle taken from the detection catalogues. The half-light radius is scaled according to the ratio of the pixel scales between Subaru and PAUS images. We keep the flux (and thus magnitude) of the galaxies constant with a high S/N to facilitate \textsc{SE} photometry. An example of a Subaru image, detection image generated from it and overlapping PAUS image are shown in figure~\ref{fig:tortorelli_fig4}. The photometric measurement is then performed on the `measurement' image, which can be either a PAUS real image or a simulated one. The forced photometry allows us to also obtain consistent colour information since the photometry is done on the same objects with the same aperture (rescaled according to the PSF) throughout the whole wavelength range.

\begin{figure}[t!]
\centering
\includegraphics[width=4cm]{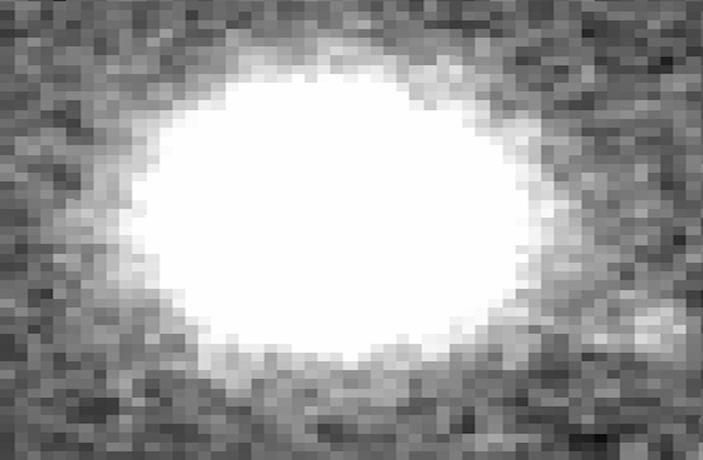}
\includegraphics[width=4cm]{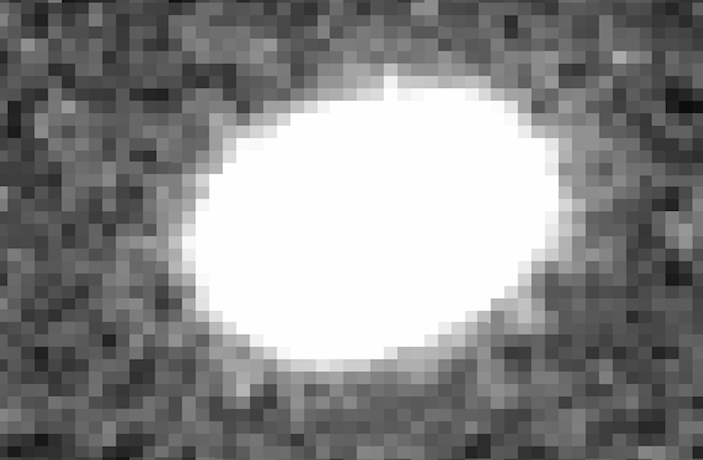} \\
\includegraphics[width=4cm]{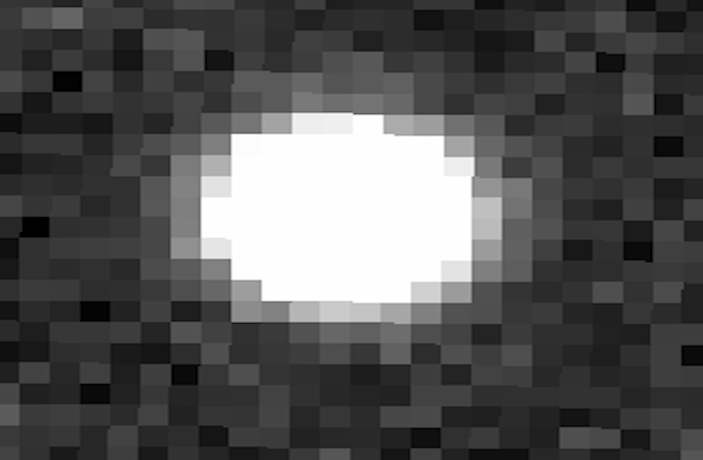}
\includegraphics[width=4cm]{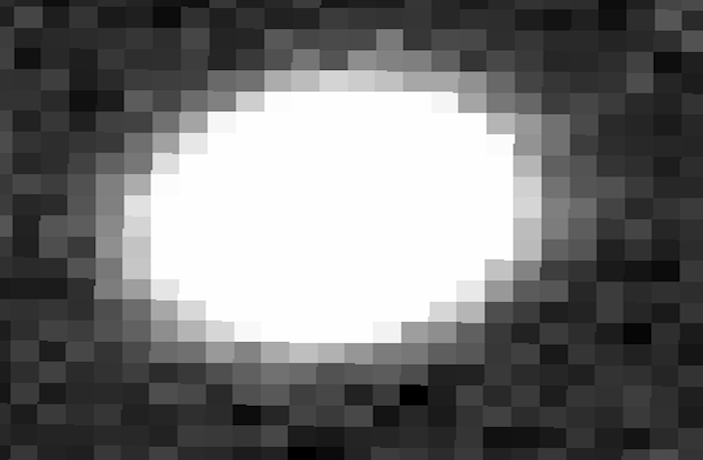} \\
\includegraphics[width=4cm]{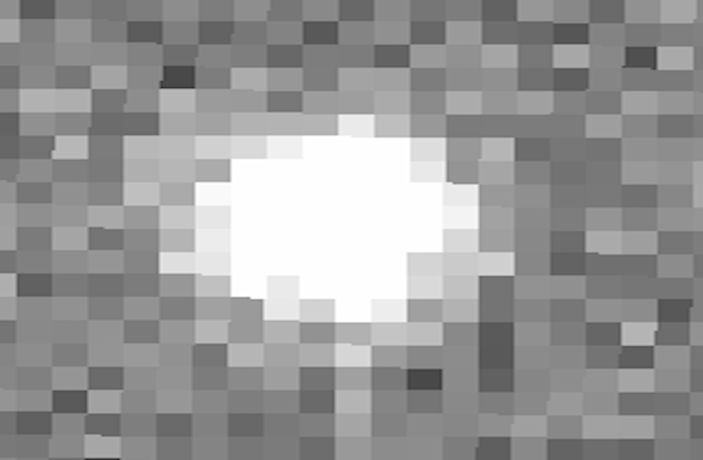}
\includegraphics[width=4cm]{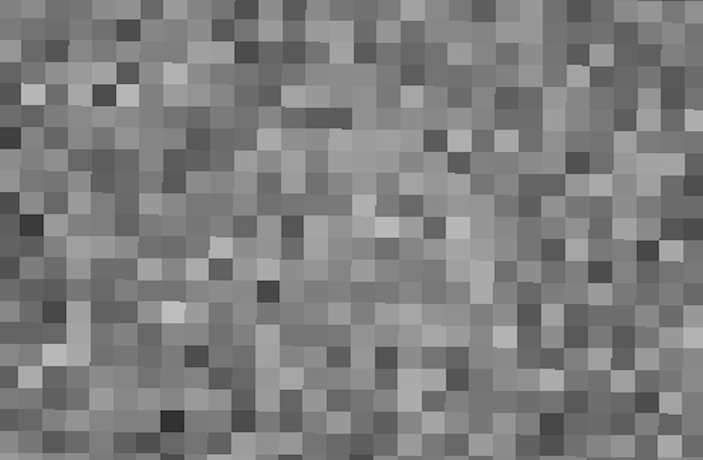} \\
\includegraphics[width=4cm]{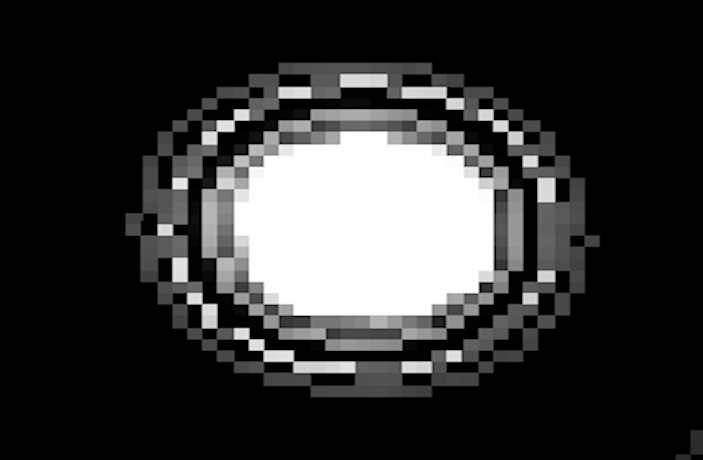}
\includegraphics[width=4cm]{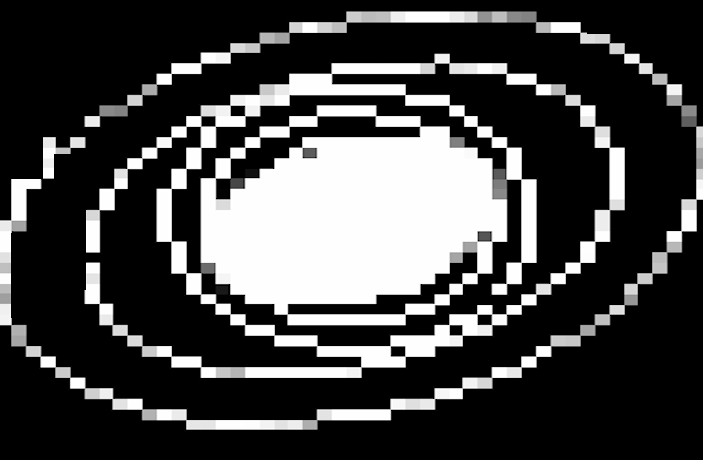}
\caption{Comparison of different S/N regimes for the pipeline. The left panels refer to a high S/N case, while the right panels to a low S/N case. The top panels show cut-outs of two different sources found in the same Subaru image. The second row shows cut-outs of the detection image, created using photometric parameters measured with the Subaru image. The third row shows the corresponding PAUS image of the two sources, which is low S/N on the right. The bottom panels show the corresponding \textsc{SE} aperture images, the sizes of which are determined on the detection images.}
\label{fig:tortorelli_fig5}
\end{figure}

In figure~\ref{fig:tortorelli_fig5}, we show an example of how \textsc{SE} behaves in two different S/N regimes. The top panels show cut-outs for two sources found in the same Subaru image. From the photometric parameters measured on the Subaru images, we create a detection image. The second row shows cut-outs of the detection image for the two sources. As explained above, these are rendered with high S/N to allow the precise determination of positions and apertures. The third row shows the PAUS image of the two sources. The left panel contains a high S/N source that can be detected by \textsc{SE} also in the single image mode. On the contrary, the right panel is dominated by background noise. The bottom panels show the corresponding \textsc{SE} aperture images, the sizes of which are determined on the detection images. The forced photometry allows us to extract information also for sources which are not directly visible on PAUS images, being below the background noise level.

\subsection{Final Catalogue Generation}

The output of our analysis on PAUS data leads to 40 catalogues (one per band). Since in the current implementation, we analysed real data and only one simulation from a specific functioning point, we produce a total of 80 catalogues per PAUS field. We then apply a cut to each catalogue by removing sources that lie less than 200 pixels from the edges of each image to avoid contamination from the scattered light we have observed coming in the filter borders from the grid of the tray \cite{tonello18}. After this cut, each catalogue contains photometric information for nearly 5000 sources. To build the final catalogue containing all 40 NB filters photometric information of each source, we need to find the overlapping images for each given PAUS image. Therefore, we create a master table with all the measured sources and then we match them. The final catalogue contains all the photometric information on the 40 NB filters for each detected source.

\section{Diagnostics}
\label{section:diagnostics}

\begin{figure}[t!]
\centering
\includegraphics[width=12cm]{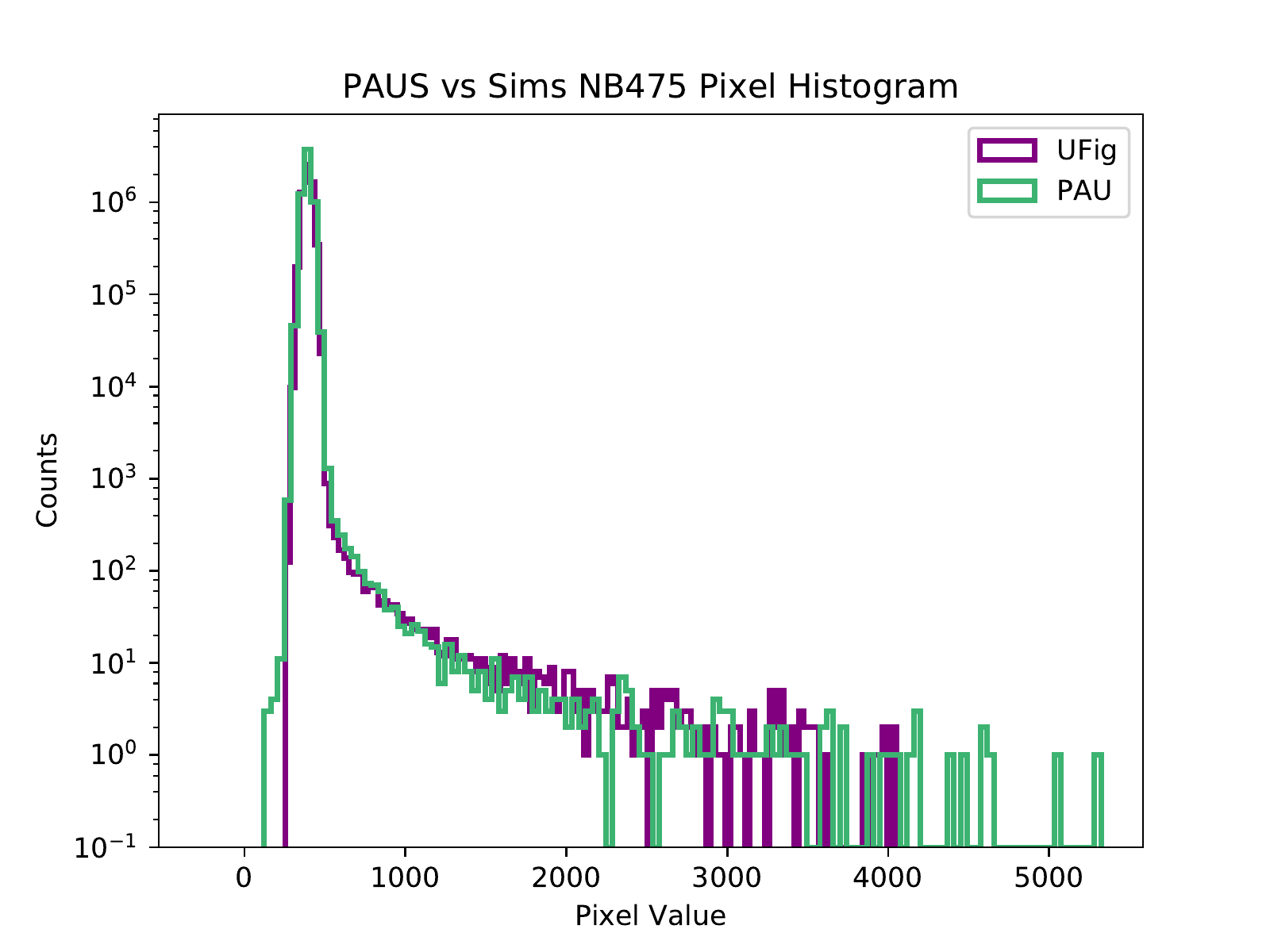}
\caption{Pixel value histograms of observed PAUS images (green line) and simulated PAUS images (purple line) are shown. On the x-axis we have the pixel value, while y-axis the corresponding counts per bin of pixel value. The two distributions statistically agree in terms of background noise level and overall shape and range of the pixel value distribution.}
\label{fig:tortorelli_fig6}
\end{figure}

To check the robustness of our method (see section~\ref{section:methodology}), we compare different diagnostics to check whether the observed and simulated PAUS images and catalogues agree statistically.

Figure~\ref{fig:tortorelli_fig6} shows a comparison of the pixel histograms of observed (green line) and simulated (purple line) PAUS images in the NB475 filter, centred at 475 nm. The first statistical property that needs to be compared is whether the background noise distributions agree. This can be seen by looking at the peak of the distributions at low pixel counts. They are in good agreement both in terms of position of the peak and width of the distribution. Then, the overall distribution at higher pixel values (which is due to galaxy and stars) needs to be compared. Also in this case, the two distribution of observed and simulated pixel counts agree well, both in terms of shape and in terms of range of pixel values.

\begin{figure}
\centering
\includegraphics[width=7.6cm]{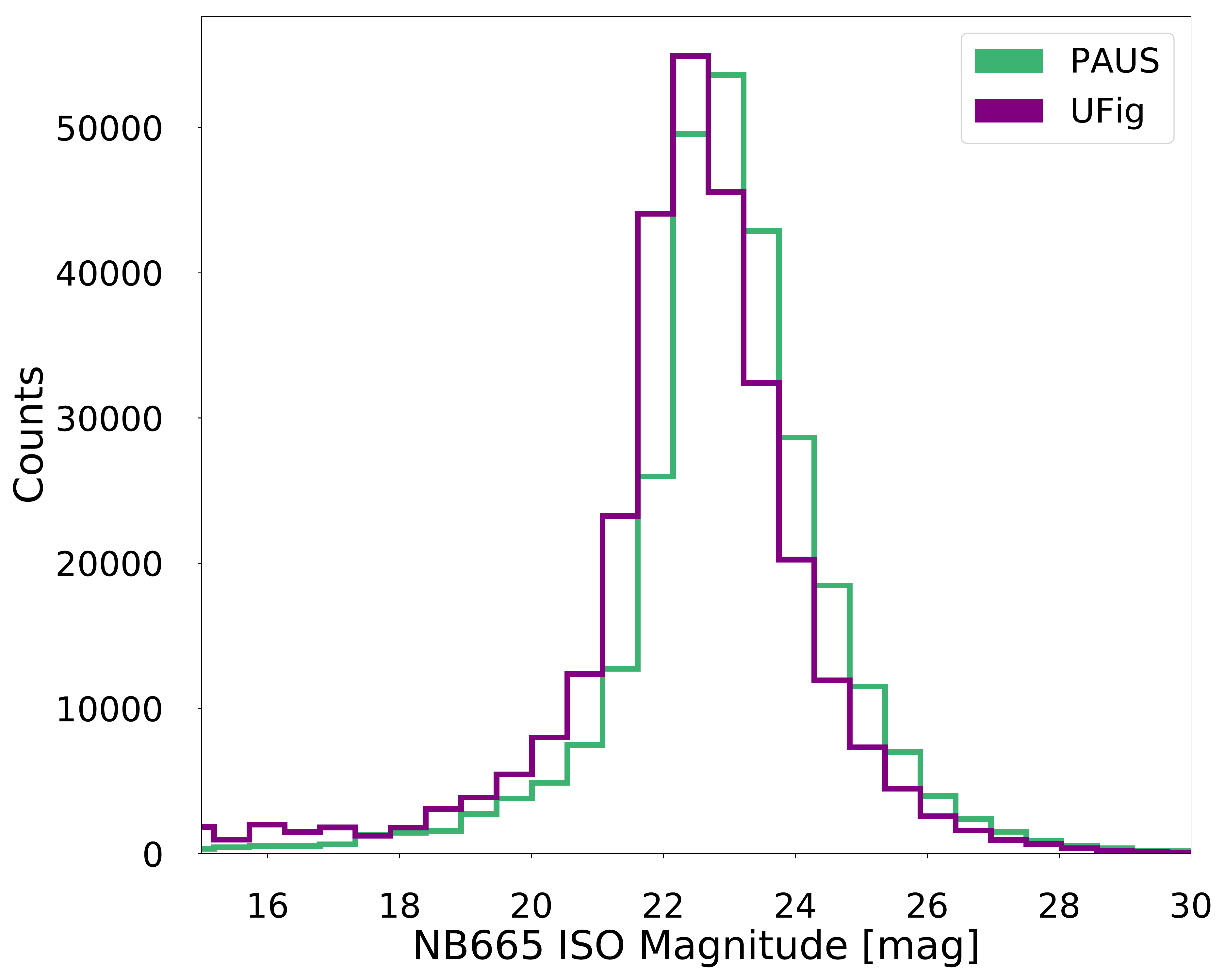}
\includegraphics[width=7.6cm]{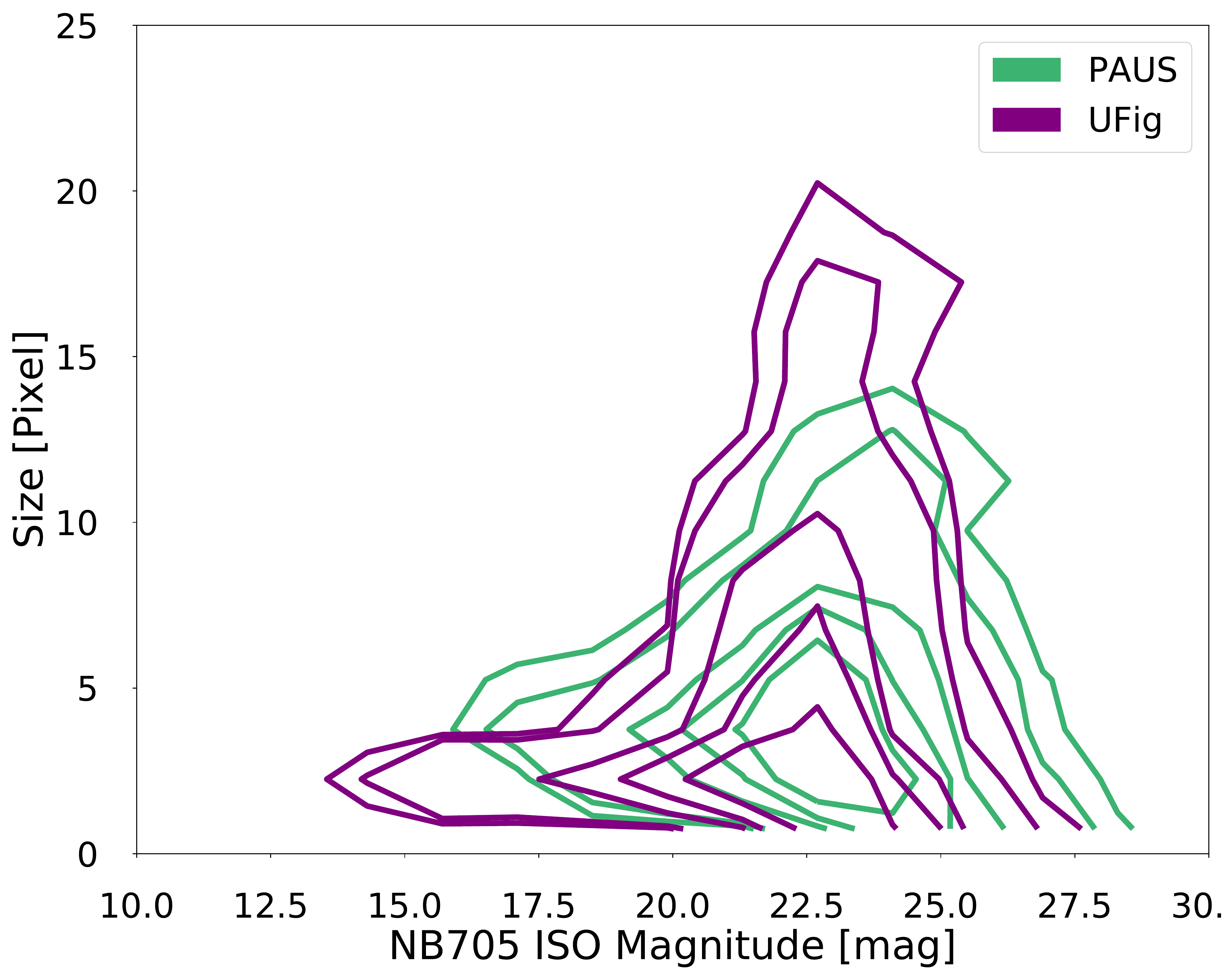}
\includegraphics[width=7.6cm]{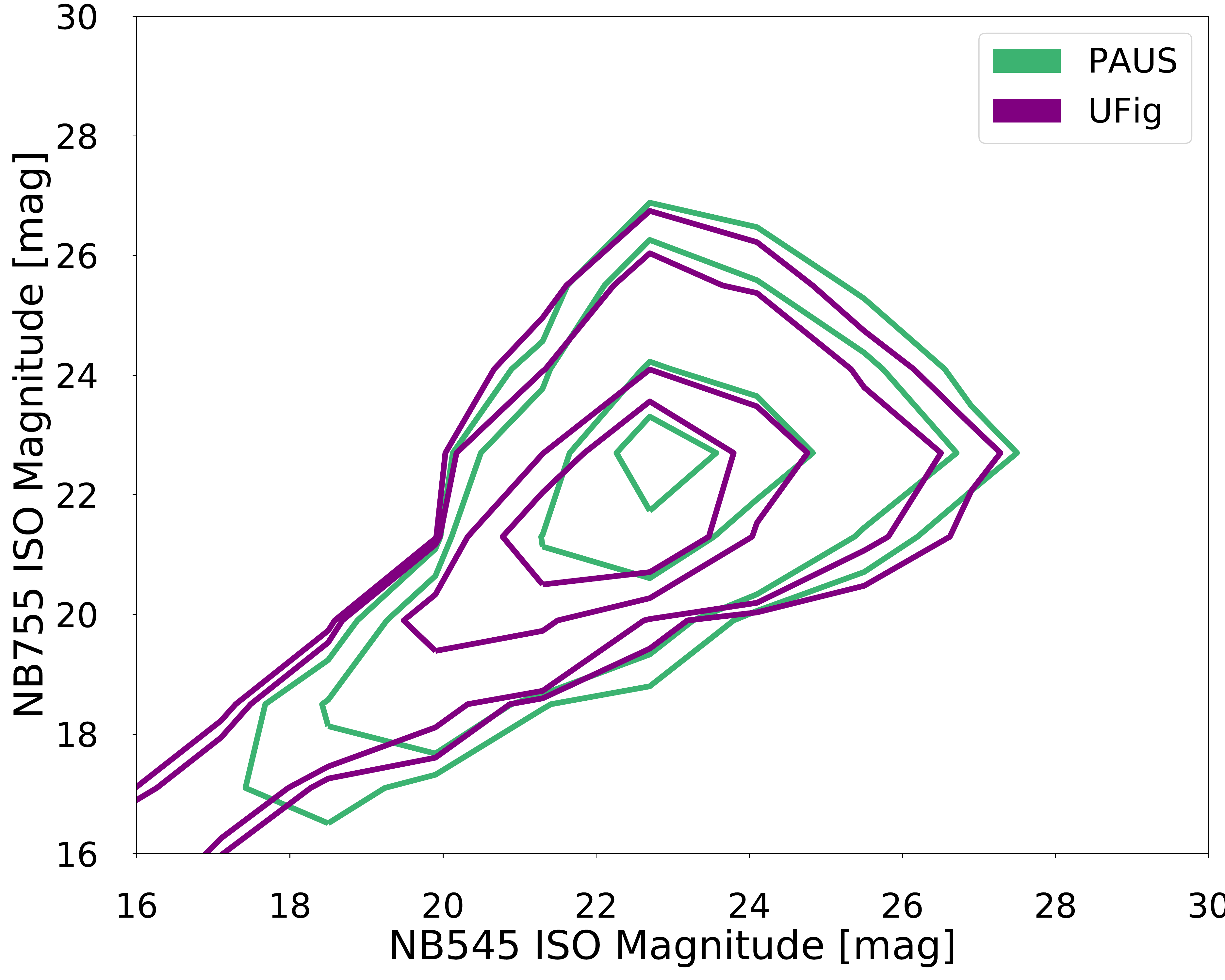}
\includegraphics[width=7.6cm]{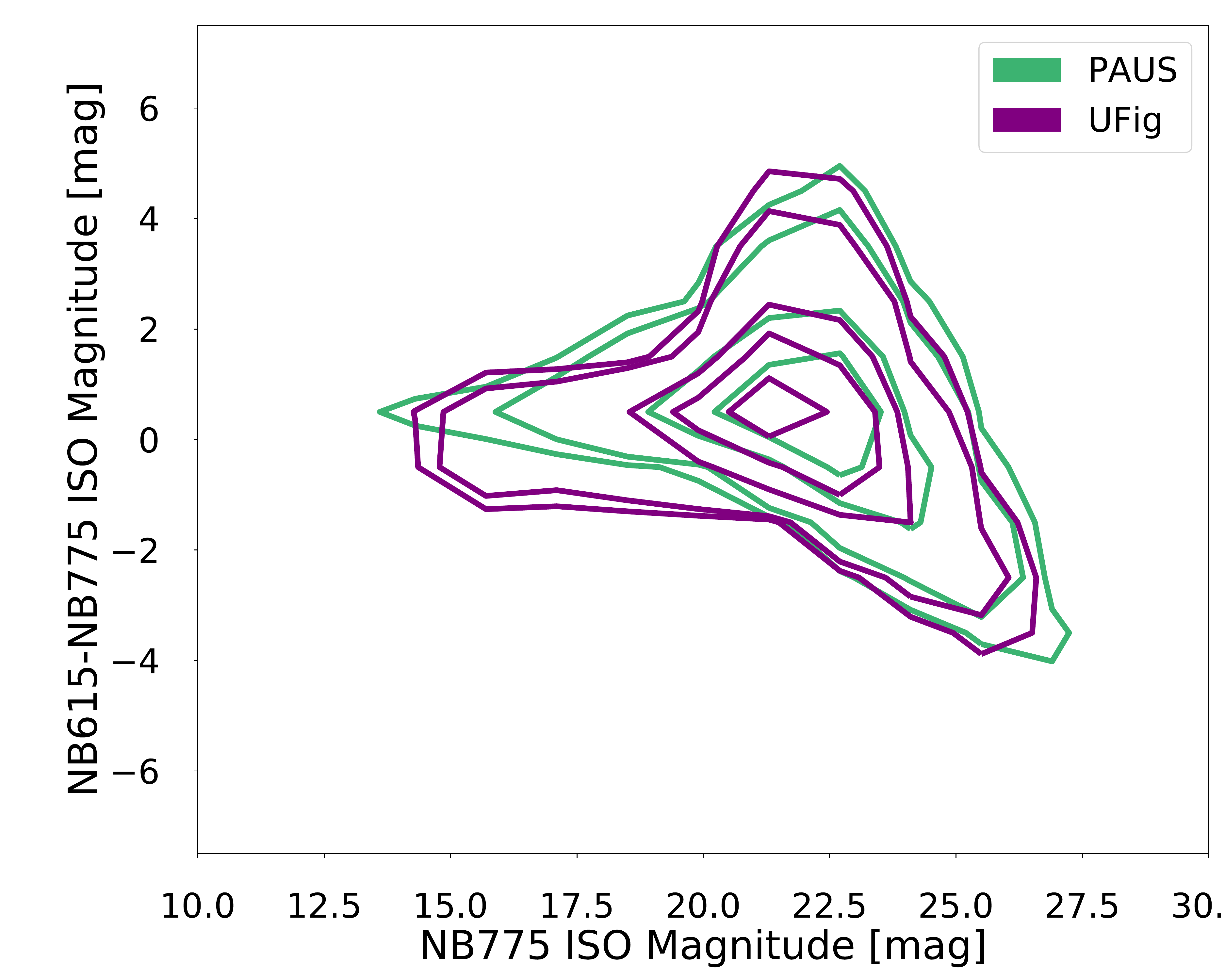}
\caption{Diagnostic plots at the image and catalogue level. The top left, top right, bottom left, bottom right panels show the magnitude histogram in the NB665 filter, the magnitude-size in the NB705 filter, the magnitude-magnitude (in the NB545 and NB755 filters) and the colour-magnitude (m$_{\mathrm{NB615}}$ - m$_{\mathrm{NB775}}$ vs m$_{\mathrm{NB775}}$) plots, respectively. Green lines refer to the real PAUS sources, while purple lines to simulated sources.}
\label{fig:tortorelli_fig7}
\end{figure}

We then compare data and simulations at the catalogue level, considering both single band and inter-band correlations. We compare the magnitude distributions for each of the 40 NB filters of our sample of PAUS fields analysed. Hereafter, unless otherwise specified, magnitudes are the `ISO' magnitudes from \textsc{SE} (see \cite{bertin96} for the definition). The top left panel of figure~\ref{fig:tortorelli_fig7} shows an example of this comparison for the NB665 filter. The two magnitude distributions (green line for observed PAUS sources and purple line for simulated ones) agree well both in terms of mean value and width. The difference in terms of detected sources, which is typically of the order of about 5 \%, is due to the difference of detected sources between observed and simulated Subaru data. This, in turns, is due to the `FLAGS' < 4 cut that has a more pronounced effect on observed Subaru data because of bad pixels due to saturated stars.

\begin{figure}
\centering
\includegraphics[width=7.5cm]{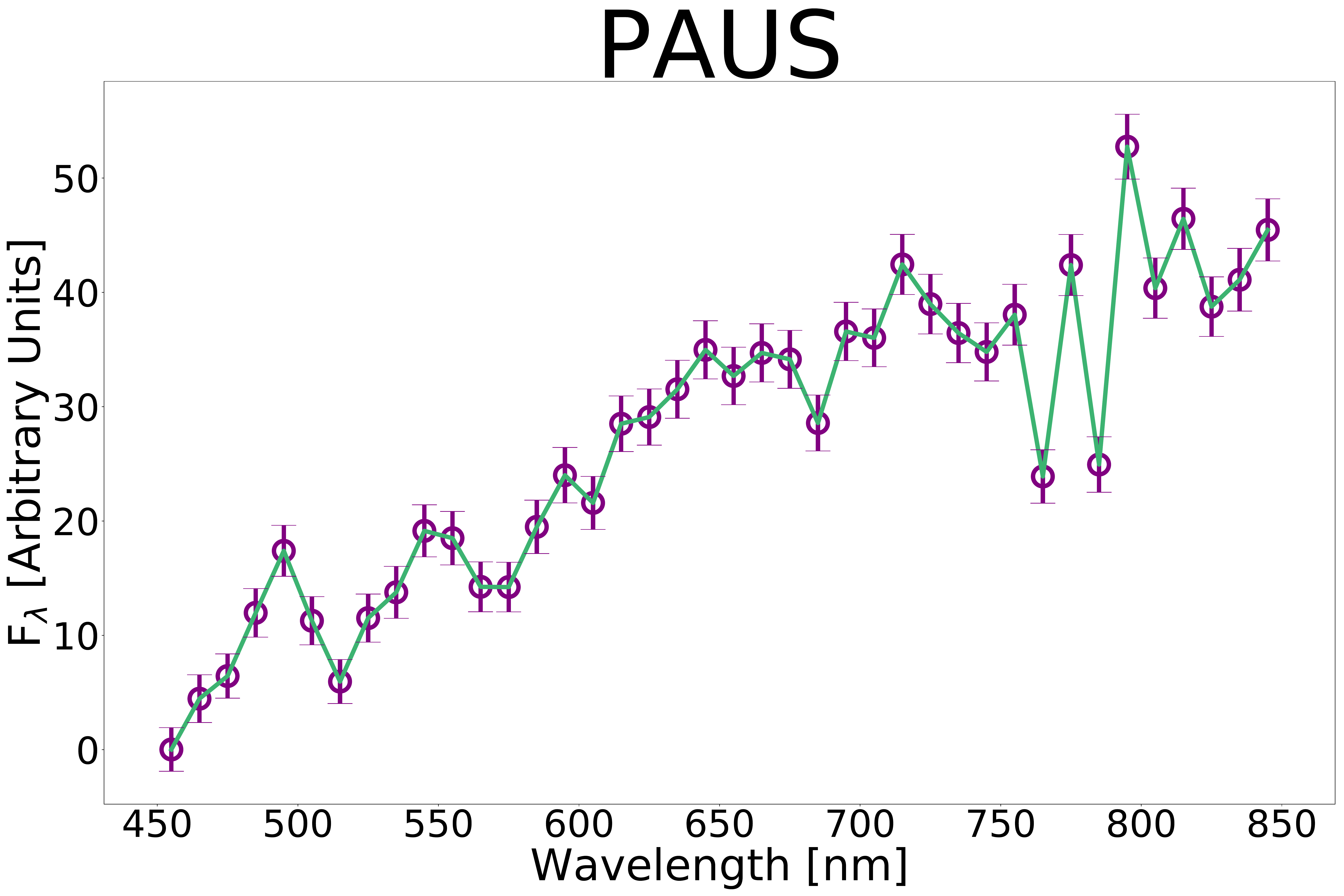}
\includegraphics[width=7.5cm]{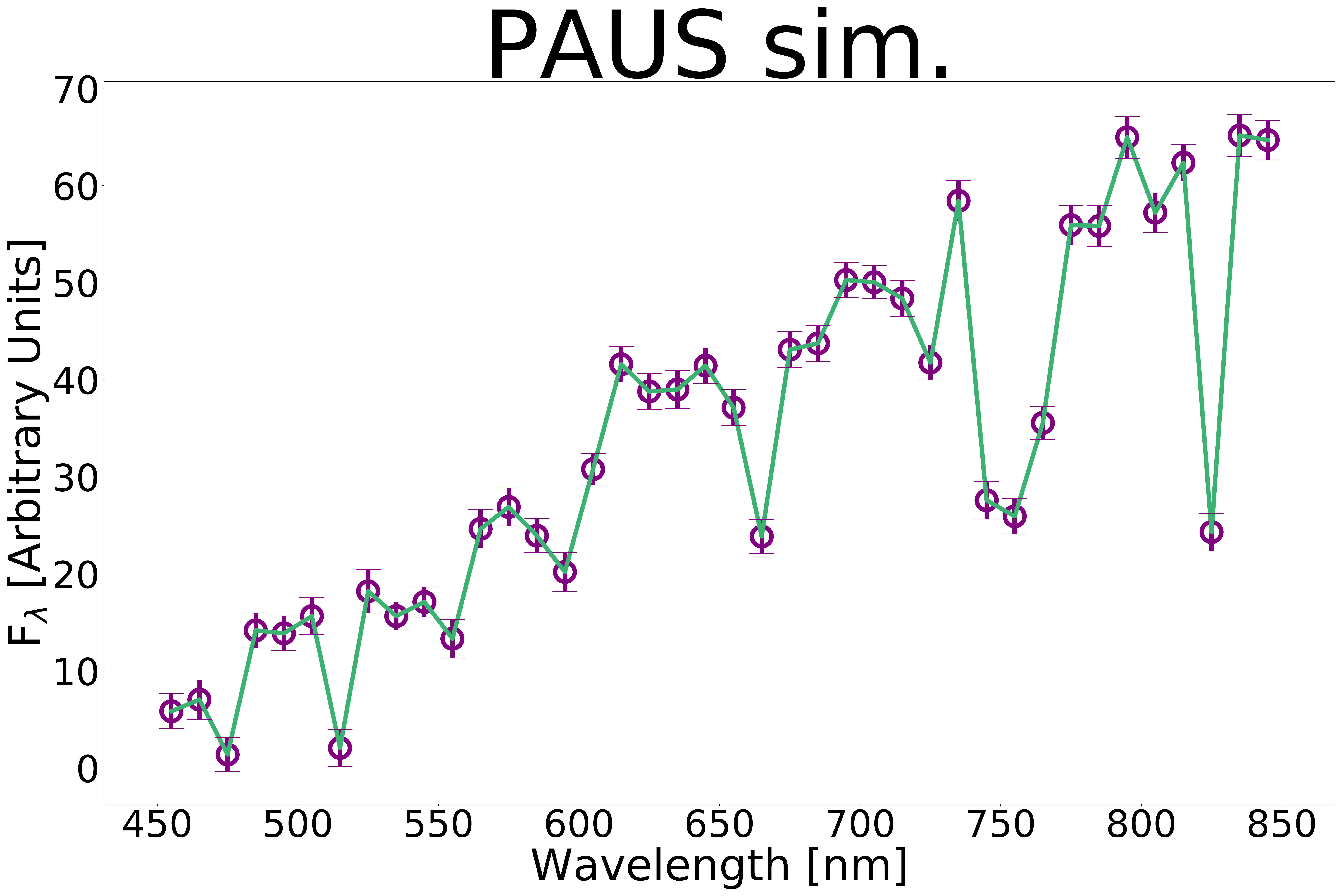}
\includegraphics[width=7.5cm]{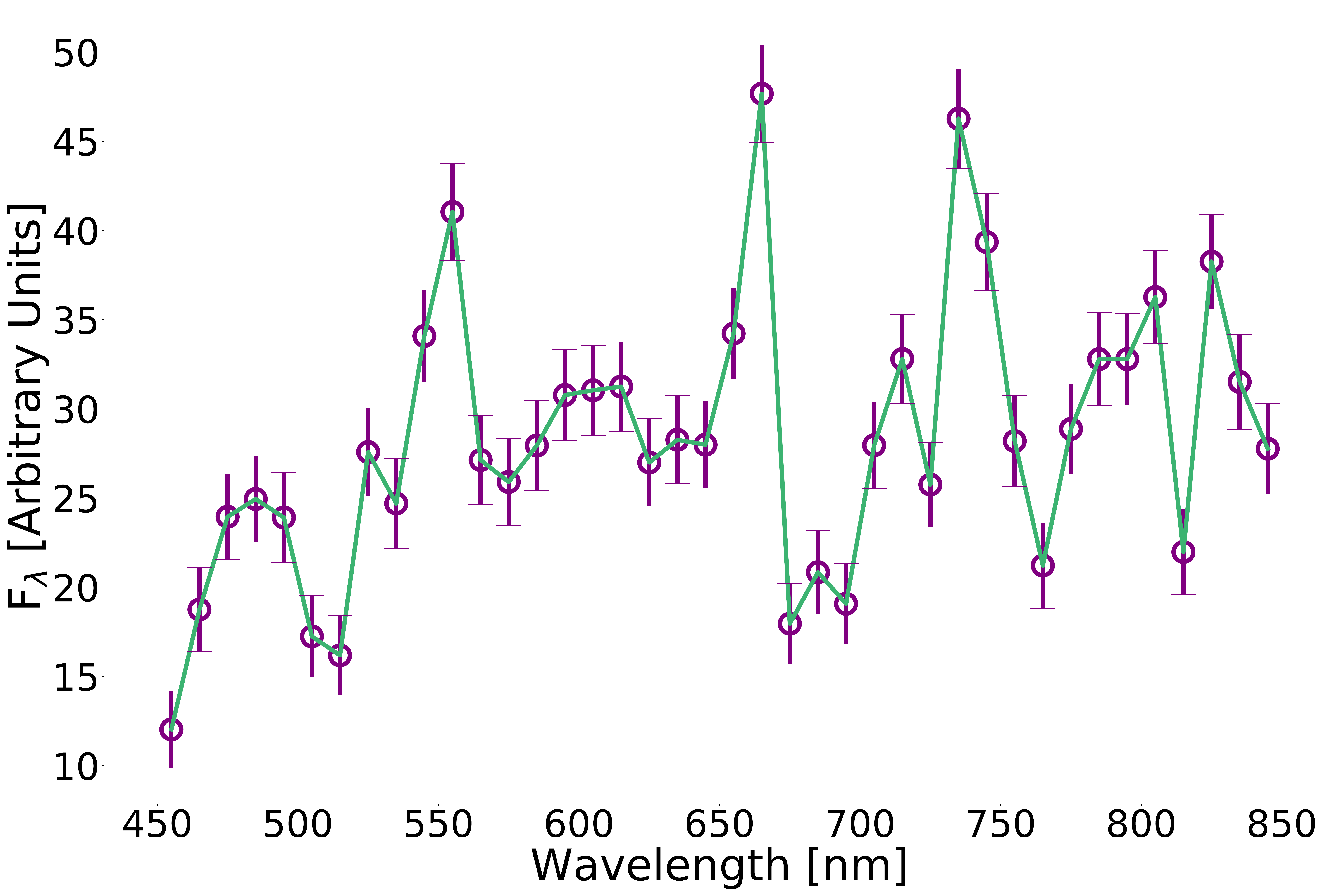}
\includegraphics[width=7.5cm]{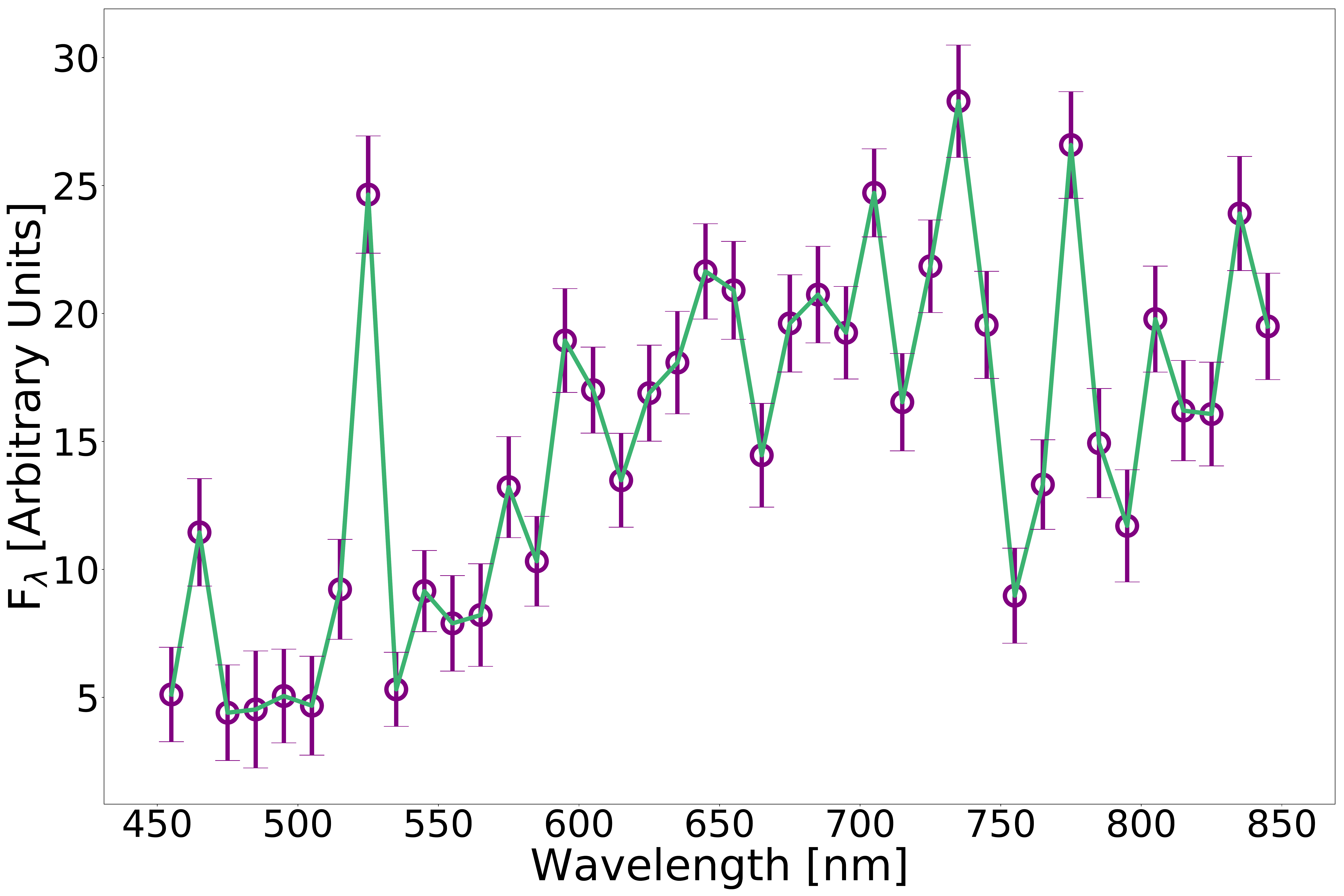}
\caption{Example NB spectra. The x-axis shows the wavelength in nanometers, while the y-axis the isophotal flux in units of counts from \textsc{SE}. The left panels show spectra obtained with 40 NB filters information from real PAUS sources. The right panels, instead, show spectra obtained from simulated PAUS sources. The top panels spectra are randomly drawn from the distribution of red galaxies, while the bottom panels spectra from the distribution of blue galaxies.}
\label{fig:tortorelli_fig8}
\end{figure}

We then compare another single band diagnostic, namely the magnitude-size relation. This has also been done for each of the 40 NB filters of our sample of PAUS fields analysed. The agreement between the observed (green points) and simulated (purple points) magnitude-size relation is good (top right panel of figure~\ref{fig:tortorelli_fig7}). It is important to stress that we are also able to fully reproduce the fixed size locus, which is the one associated to stars in the magnitude-size plot, highlighting the goodness of our image simulator in also simulating stars.

We then consider also inter-band correlations (bottom panels in figure~\ref{fig:tortorelli_fig7}). The bottom left panel shows an example of magnitude-magnitude distribution in two NB filters, namely NB545 (x-axis) and NB755 (y-axis). The bottom right panel shows as an example the colour-magnitude distribution for the colour m$_{\mathrm{NB615}}$ - m$_{\mathrm{NB775}}$ and the magnitude in the NB775 filter. Except for stars, which are easily recognisable as objects with zero colour in the current UFig runs, in both cases, observed (green points) and simulated (purple points) distributions agree well.

We show in figure~\ref{fig:tortorelli_fig8} example spectra built with the 40 NB filters information from both data (left panels) and simulations (right panels). Fluxes from 40 NB filters in the wavelength range 450--850 nm gives a spectrum with a spectral resolution of R = $\lambda / \Delta \lambda$ = 50 at 650 nm. This resolution allows us to distinguish features in spectra such as those of absorption and emission lines. The different spectra show the main type of galaxies we find in PAUS data. We separate red from blue galaxies using the SDSS colour cut in \cite{eisenstein01} and we randomly draw spectra from these two distributions. The top panels show spectra belonging to the red galaxy population, while the bottom panels to objects belonging to the blue sample. It is also important to see how these two populations are well represented in UFig simulated spectra (right panels).

\section{Results}
\label{section:results}

\begin{figure}
\centering
\includegraphics[width=16cm]{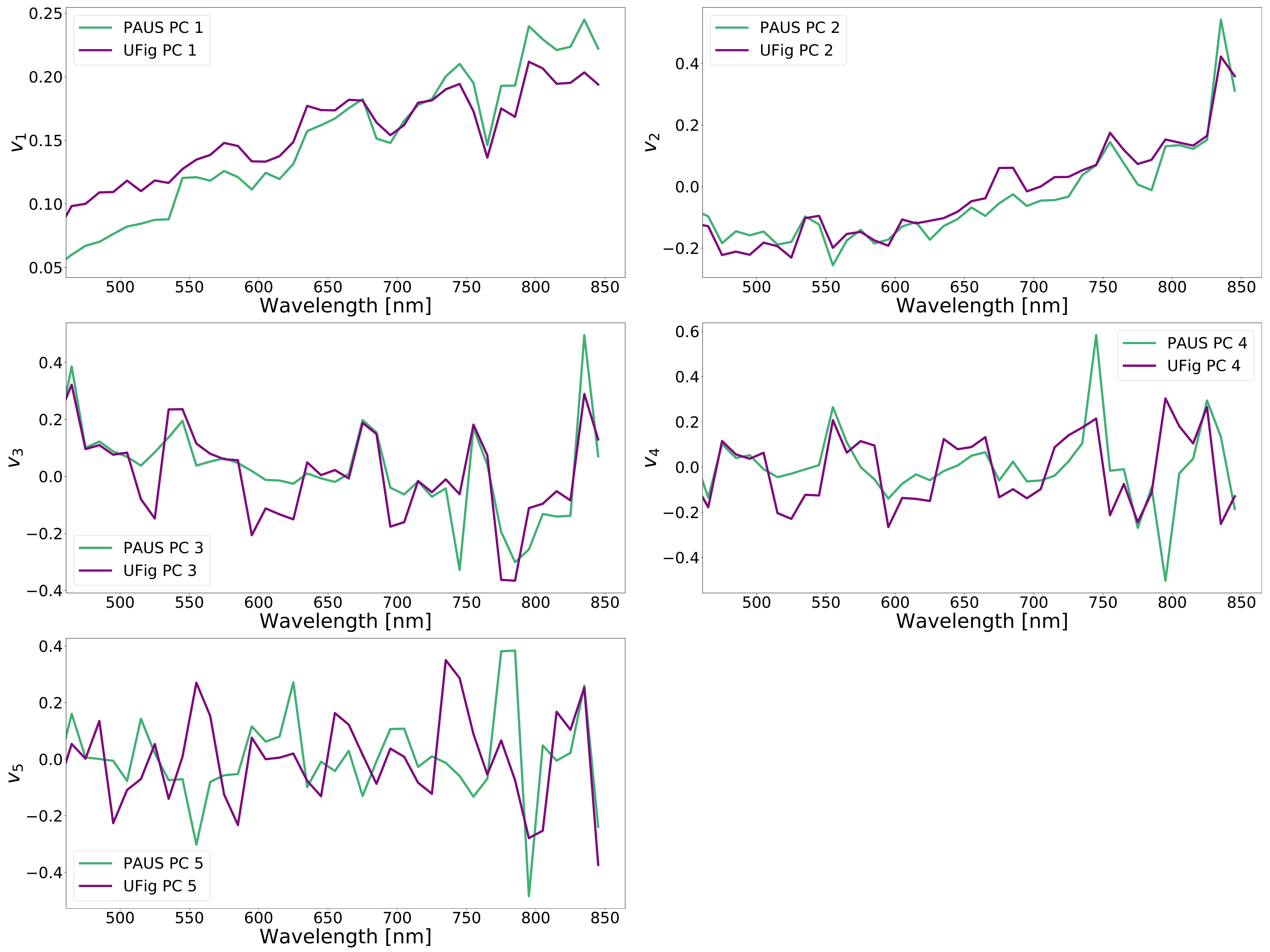}
\caption{From top left to bottom left panels, the first 5 real and simulated PCs are compared. Green lines represent observed PAUS PCs, while purple lines represent simulated ones. PCs have been normalized. The x-axis shows the wavelength in nm, while the y-axis shows the normalized PC value.}
\label{fig:tortorelli_fig9}
\end{figure}

The PAUS dataset contains in principle more spectral information compared to a photometric dataset based only on broad-bands as in~\cite{herbel17}. The usual single band or inter-band correlations do not capture all the available information and with 40 NB filters even visualizing them is difficult. Therefore we need a diagnostic to capture the global information from such a high dimensional dataset. There are several ways of tackling this type of problem and we decided to explore a PCA based approach. In this section, first we perform PCA on the galaxy catalogue obtained with our pipeline and we compare the decomposition of the derived principal components (PCs) on data and simulations. Then, we select Luminous Red galaxies (LRG) at z > 0.4 through cuts in colour space and we perform PCA on this sample. This is motivated by the fact that one of the goal of the PAU Survey is to obtain photo-z resolution on LRGs roughly one order of magnitude better than current state-of-the-art photometric surveys \cite{padilla16} and by the fact that the cut at z > 0.4 is magnitude independent.

\subsection{PCA}
\label{pcasubsection}

\begin{figure}
\centering
\includegraphics[width=15cm]{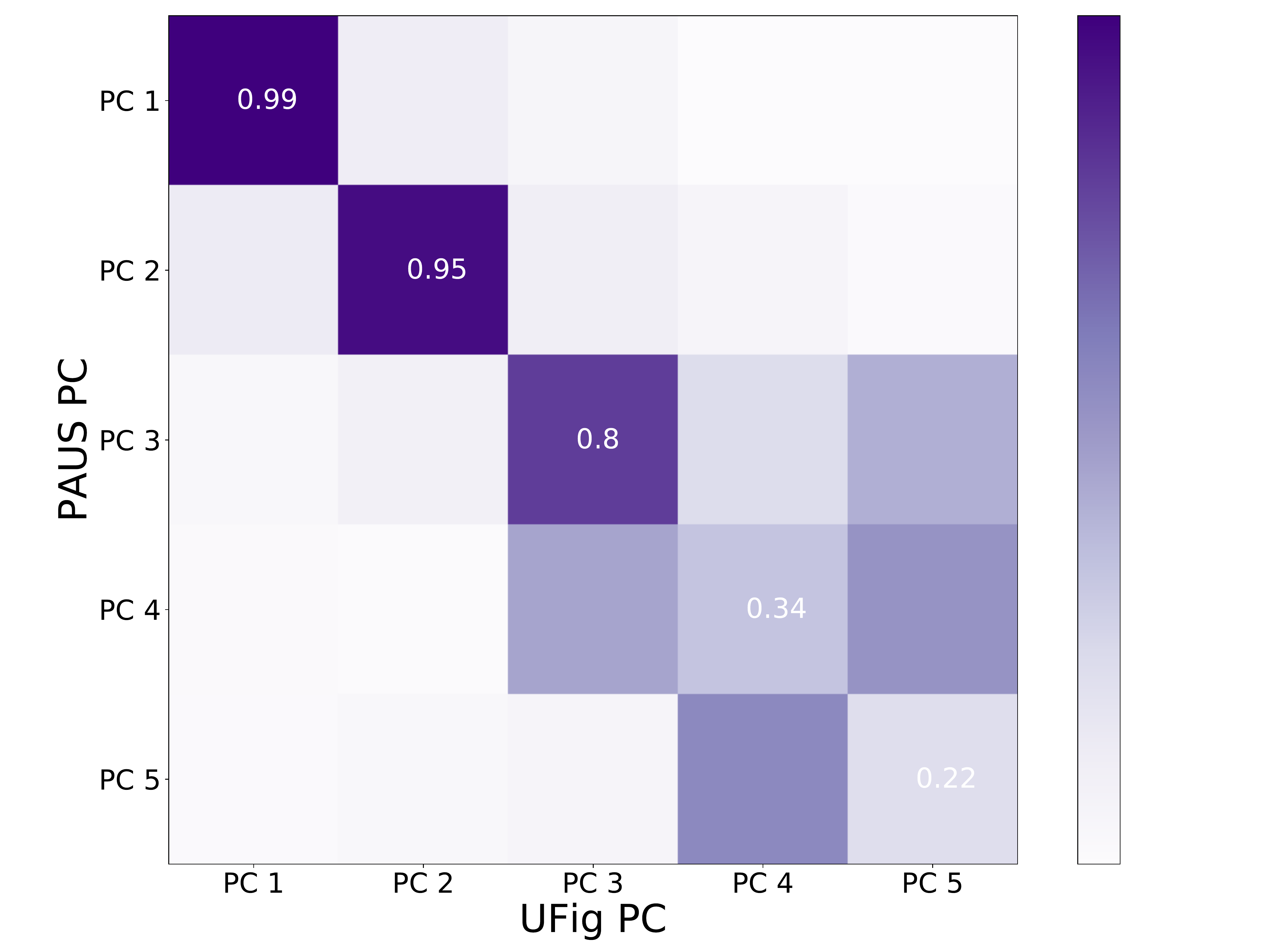}
\caption{`Mixing' matrix between observed and simulated PCs. The x-axis shows the simulated PCs, while the y-axis shows the observed ones. Being orthonormal, the colour range goes from 0 to 1.}
\label{fig:tortorelli_fig10}
\end{figure}

PCA is a way to capture the information from a noisy dataset by decomposing it into a basis set (see~\cite{tutorialpca} for a review on PCA). The purpose of PCA is to find the direction in a noisy high-dimensional dataset that capture most of the information, i.e. the direction that has the highest variance. Therefore, the first PC is the one with the highest variance and the following ones have to be orthonormal to each other with decreasing values of the variance.

\begin{figure}
\centering
\includegraphics[width=16cm]{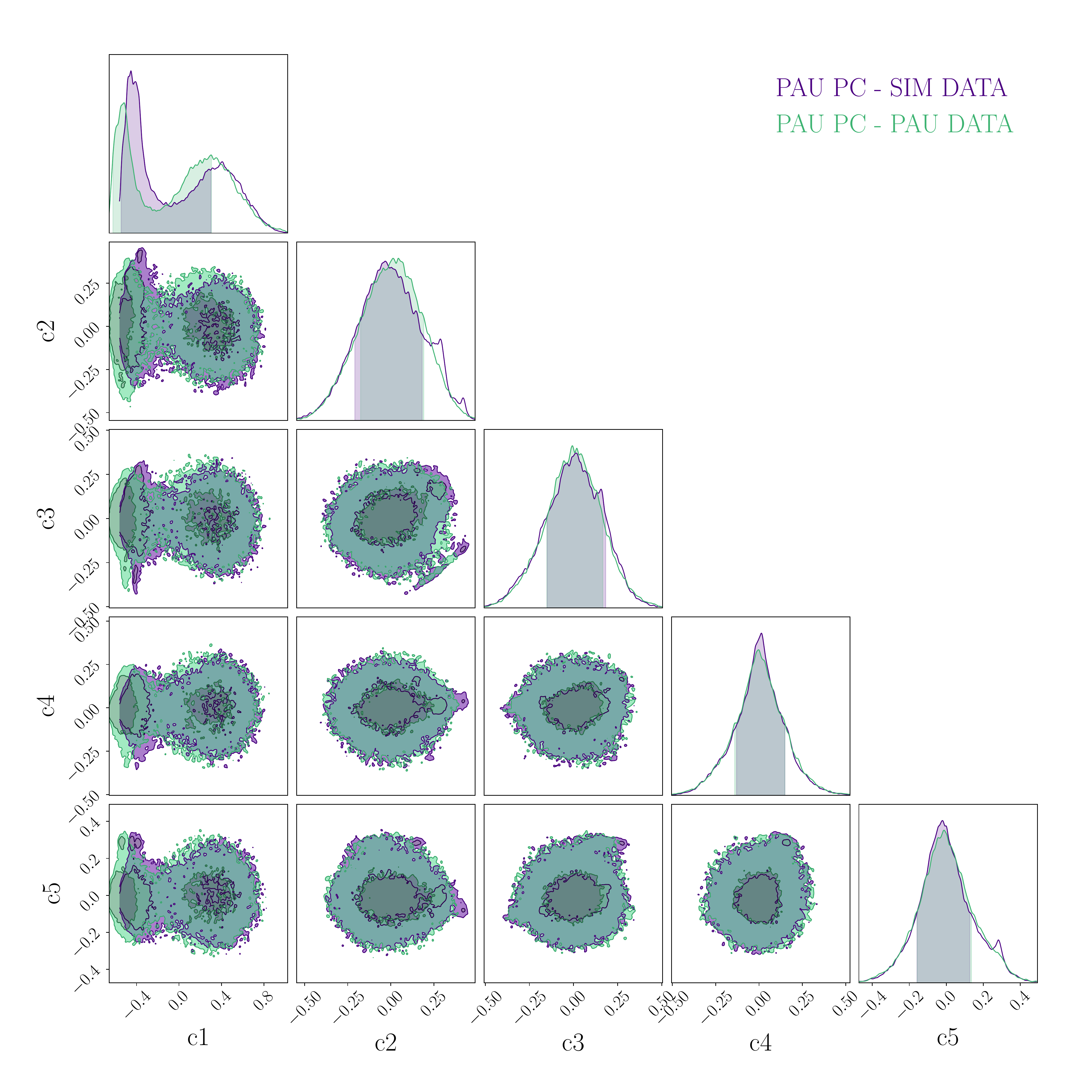}
\caption{Corner plot of the different coefficient distributions. The coefficients are built through the scalar product between the observed PAUS PCs and observed data (green contours and histograms) and between the observed PAUS PCs and simulated data (purple contours and histograms). `c' is short for coefficient.}
\label{fig:tortorelli_fig11}
\end{figure}

\begin{figure}
\centering
\includegraphics[width=16cm]{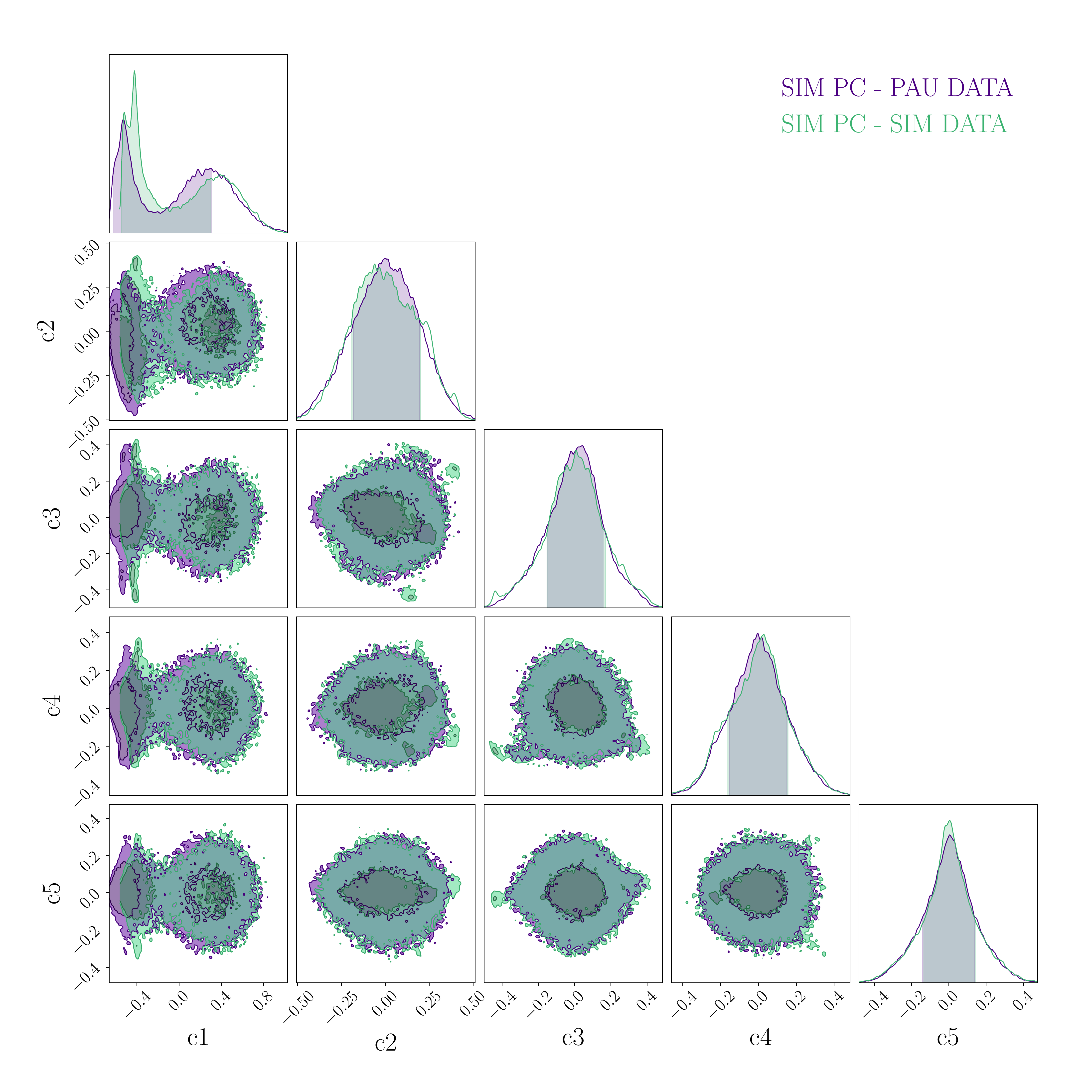}
\caption{Corner plot of the different coefficient distributions. The coefficients are built through the scalar product between the simulated PAUS PCs and real data (purple contours and histograms) and between the simulated PAUS PCs and simulated data (green contours and histograms). `c' is short for coefficient.}
\label{fig:tortorelli_fig12}
\end{figure}

We perform PCA with the singular value decomposition technique on the catalogues generated via forced photometry, both for data and simulations. Given the PCs, data and simulations can be decomposed as:
\begin{equation}
\begin{split}
\mathrm{f}(\lambda) &= \sum_{\mathrm{j}} \mathrm{a}_{\mathrm{j},\mathrm{data}} \phi_{\mathrm{j}}(\lambda) \\
\mathrm{f}^{'}(\lambda) &= \sum_{\mathrm{j}} \mathrm{b}_{\mathrm{j},\mathrm{sims}} \psi_{\mathrm{j}}(\lambda)
\end{split}
\label{equa}
\end{equation}
where $\mathrm{f}(\lambda)$ and $\mathrm{f}^{'}(\lambda)$ represent the real and simulated datasets, $\mathrm{a}_{j,\mathrm{data}}$ and $\mathrm{b}_{j,\mathrm{sims}}$ are the decomposition coefficients of the dataset in terms of PCs, and $\phi_i(\lambda)$ and $\psi_i(\lambda)$ are the real and simulated PCs, respectively. Fluxes are mean subtracted and normalized to unit norm for each object in the catalogues. We separate galaxies from stars with the \textsc{SE} CLASS\_STAR parameter and we perform PCA on the resulting galaxy sample ($\sim$ 46000 galaxies for both data and simulations). Figure~\ref{fig:tortorelli_fig9} shows the comparison between the first 5 real (green lines) and simulated (purple lines) PCs. We use 5 components because the spectra assigned by \textsc{UFig} to galaxies are linear combinations of 5 basis spectra taken from \textsc{kcorrect} templates (see section~\ref{section:UFig}).

\begin{figure}
\centering
\includegraphics[width=16cm]{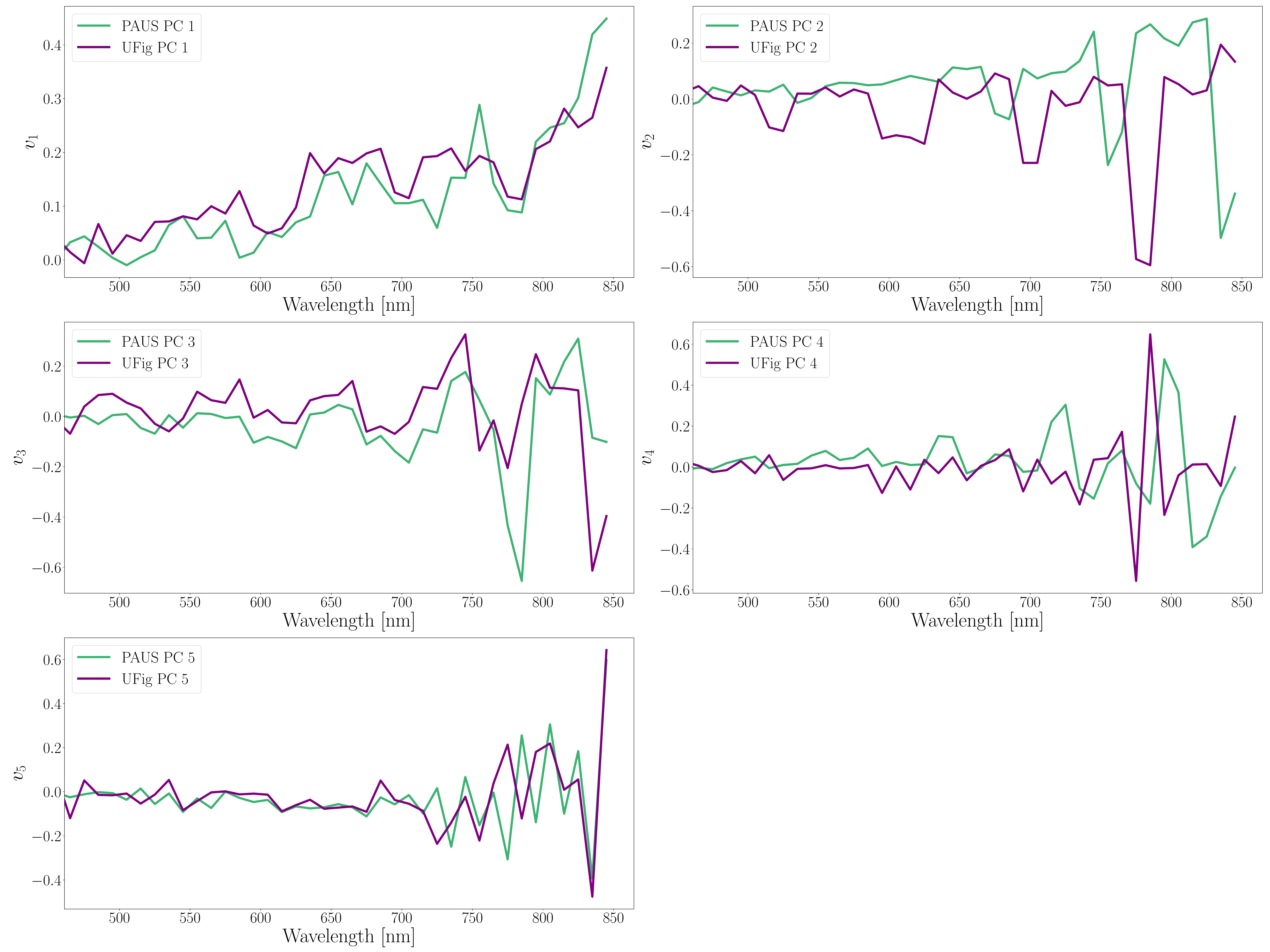}
\caption{From top left to bottom left panels, the first 5 real and simulated PCs resulting from the LRGs sample are compared. Green lines represent observed PAUS LRGs PCs, while purple lines represent simulated ones. PCs have been normalized. The x-axis shows the wavelength in nm, while the y-axis shows the normalized PC value.}
\label{fig:tortorelli_fig13}
\end{figure}

From a visual inspection, the first three PCs show good agreement, while the remaining two only moderately agree. In particular, the first two PCs capture the bulk of the physical information coming from the spectra. Individual features are not clearly visible because of the spectral resolution and because the spectra are analyzed in the observed frame. The shapes are clearly those of a population of galaxies, having a spectrum which decreases in intensity with decreasing wavelength, namely red galaxies. The third PC seems to capture, instead, a population of sources having emission lines and fluxes rising towards the blue part of the spectrum. Therefore this suggests that the third PC represents the population of galaxies characterized by recent star-formation. The fourth and fifth PCs only moderately agree between data and simulations and they are supposed to show the relevant recurring spectral features, broadened by the effect of the variety of redshifts analyzed. However, due to the lack of statistics, these features are not clearly visible in our PCA.

It is worth to notice that the UFig galaxy population in the first PC shows a flatter shape and a more pronounced flux in the bluer wavelength with respect to the PAUS observed one. It is remarkable to observe that we obtain the same results found by \cite{fagioli18}, but using a photometric dataset, which in principle contains less spectral information than spectroscopic one about the underlying galaxy population. \cite{fagioli18} justify their result to the presence of an overall bluer population of galaxies in UFig basis spectra with respect to their SDSS sample. Our results confirm that this is the case and the conclusion can be extended to the PAUS sample too. This implies that an optimization of the functioning point used in this work is required and can be achieved in a future work with an Approximate Bayesian Computation (ABC)~\cite{akeret15} run on combined spectroscopic and photometric datasets, including the PAUS full one.

A testable way to quantify the agreement between the real and simulated PCs is by using the `mixing' matrix. The idea of PCA is to construct a basis set to describe a dataset that has to be orthonormal. PCs of data and simulations are internally orthonormal, i.e.
\begin{equation}
\begin{split}
\mathrm{\int \phi_i(\lambda) \phi_j(\lambda) d\lambda} &= \delta_{\mathrm{ij}} \\
\mathrm{\int \psi_i(\lambda) \psi_j(\lambda) d\lambda} &= \delta_{\mathrm{ij}}
\end{split}
\end{equation}
but the relation to each other still needs to be tested. One way this can be quantified is through the mixed scalar product between the real and simulated PCs:
\begin{equation}
\mathrm{M_{ij} = \int \phi_i(\lambda) \psi_j(\lambda) d\lambda}
\end{equation}
The matrix of these scalar products is shown in figure~\ref{fig:tortorelli_fig10}. The x-axis represent the 5 simulated PCs and the y-axis the real ones. We report the value of the scalar product between the same order PCs along the diagonal. Ideally, one would like the matrix to be the unity matrix, such that all the components overlap between each other (the scalar products between same order PCs is 1 and the off-diagonal components are 0). This would tell us that data and simulations statistically agree and the single components for the two datasets are identical.

\begin{figure}
\centering
\includegraphics[width=16cm]{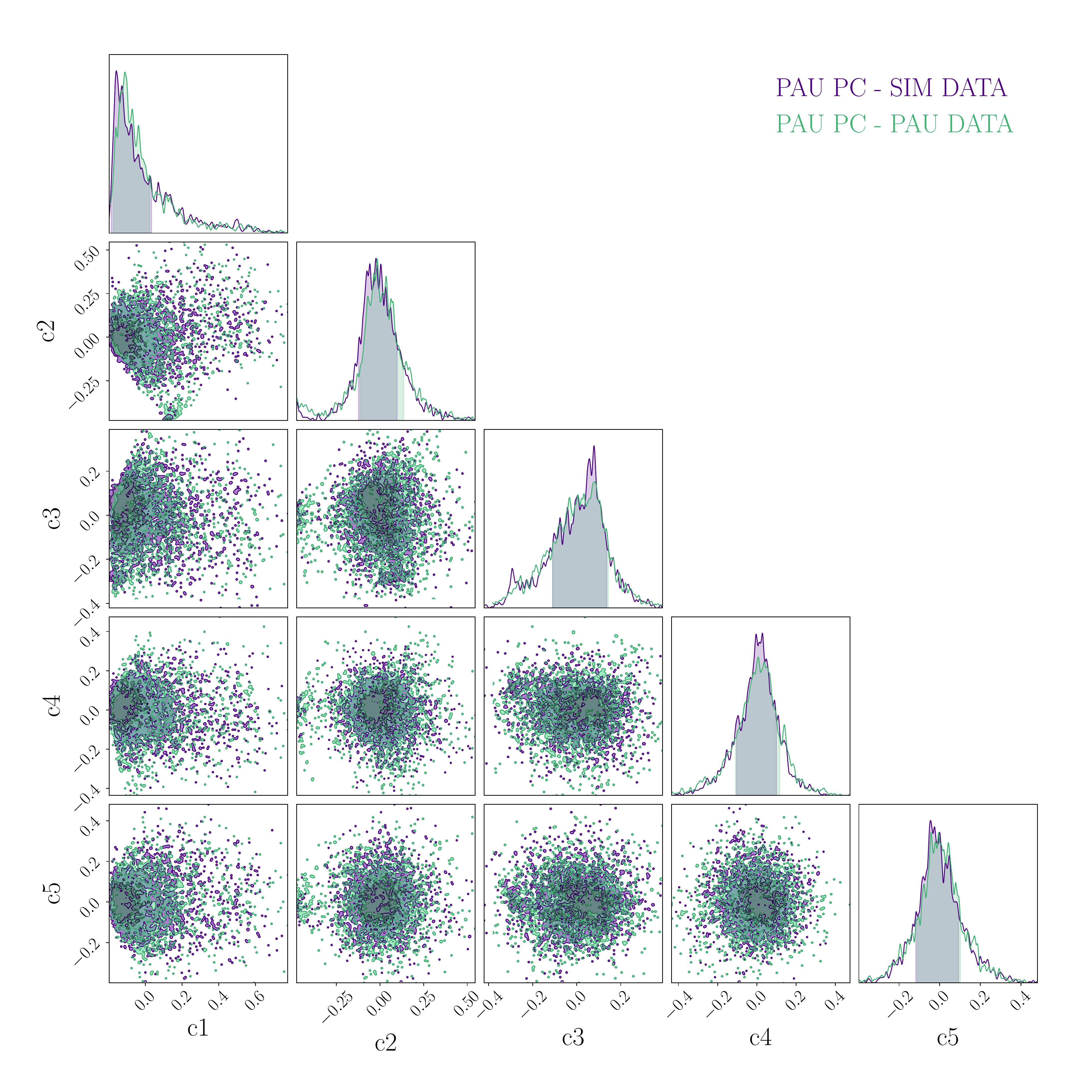}
\caption{Corner plot of the different coefficient distributions of the LRGs sample. The coefficients are built through the scalar product between the observed PAUS PCs and observed data (green contours and histograms) and between the observed PAUS PCs and simulated data (purple contours and histograms). `c' is short for coefficient.}
\label{fig:tortorelli_fig14}
\end{figure}

We find that the first three PCs agree well between data and simulations. The other two show moderate agreement (as it is already visible from figure~\ref{fig:tortorelli_fig9}), indeed the matrix has diagonal elements value of 0.34 and 0.22. The non-diagonal elements have values which are significantly smaller than the diagonal ones. This is particularly true up to the third PC.

In order to quantify how far from a unity matrix our mixing matrix is, we decide to use a distance metric given by the ratio between the diagonal elements product and the matrix determinant. In our case, the value is 
\begin{equation}
\epsilon = \left \lVert \frac{\prod(\mathrm{M_{ii}})}{\mathrm{det(M_{ij})}} \right \rVert = 0.68
\end{equation}
which is reasonably close to the value of 1 expected for a unity matrix. This value can be achieved by adjusting the parameters of the galaxy model with an ABC run in a future work.

\subsection{Decomposition Coefficients Distribution}

Another way of quantifying the agreement between the observed and simulated PCs is to look at the coefficient distributions. These are the coefficients resulting from the scalar product between the PCs of either data or simulations with the observed and simulated normalized spectra ($\mathrm{a_j}$ and $\mathrm{b_j}$ in Equations \ref{equa}). We show in figure~\ref{fig:tortorelli_fig11} and \ref{fig:tortorelli_fig12} the comparisons between the coefficient distributions built using the PCs resulting from the PAUS dataset and the PAUS simulated dataset, respectively.

The contour plots of the different component coefficient distributions show good agreement, both for PAUS real PCs and PAUS simulated PCs. A little discrepancy is however visible in the contour plots involving the first coefficient (first column of figure \ref{fig:tortorelli_fig11} and \ref{fig:tortorelli_fig12}). The discrepancy seems to be more pronounced for simulations rather than for data. In an upcoming work, the agreement of the coefficient distributions and of the PCs (see section \ref{pcasubsection}) will be improved with a fit to the model by using the ABC method on the full PAUS dataset.

\subsection{Luminous Red Galaxies Cut}

\begin{figure}
\centering
\includegraphics[width=16cm]{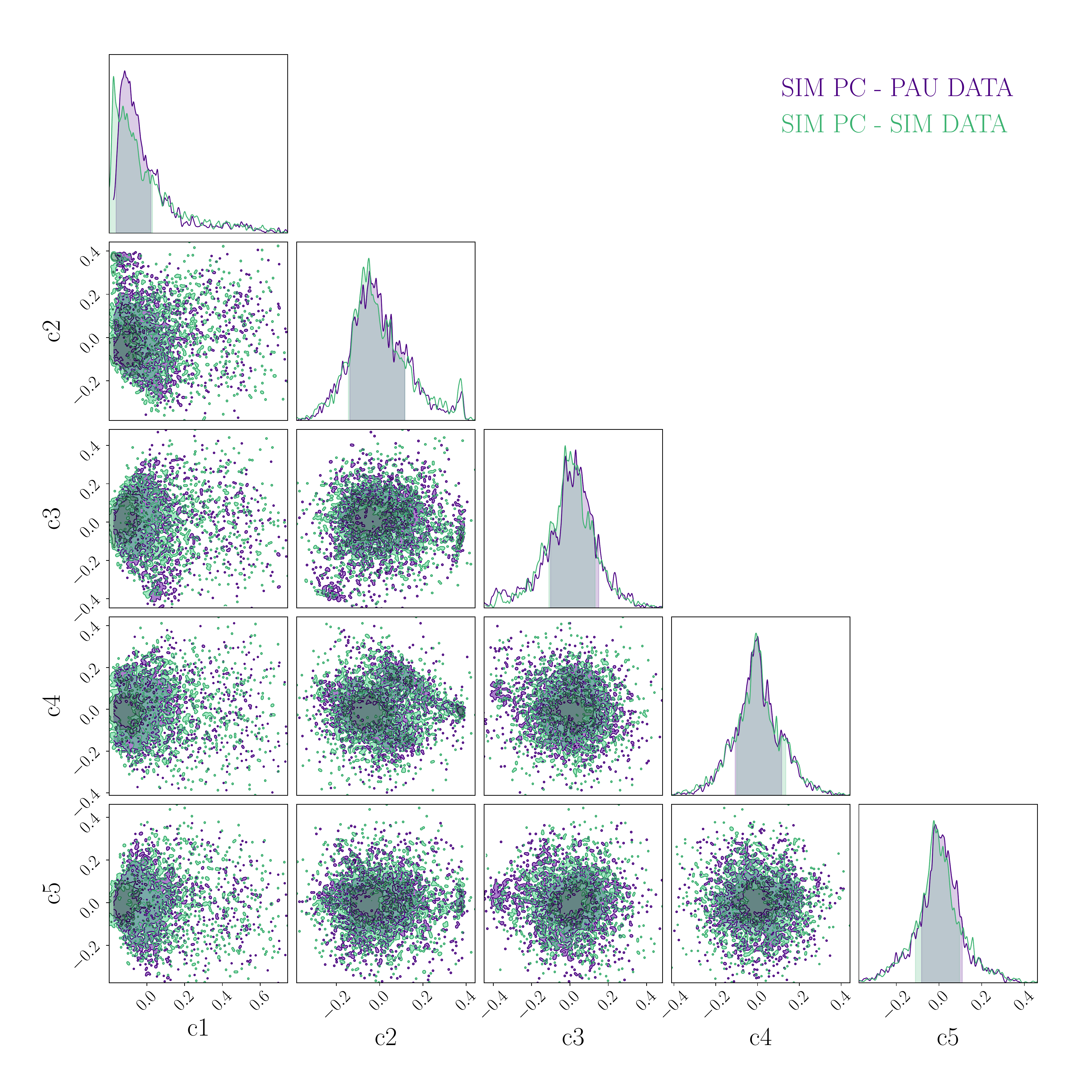}
\caption{Corner plot of the different coefficient distributions of the LRGs sample. The coefficients are built through the scalar product between the simulated PAUS PCs and real data (purple contours and histograms) and between the simulated PAUS PCs and simulated data (green contours and histograms). `c' is short for coefficient.}
\label{fig:tortorelli_fig15}
\end{figure}

As motivated in the beginning of section \ref{section:results}, we perform a cut on real and simulated final galaxy catalogues to select a sample of LRGs. We follow the prescription in \cite{eisenstein01} to select LRGs at z > 0.4. The sample consists of $\sim$ 2400 LRGs for both data and simulations. We show in figure \ref{fig:tortorelli_fig13} the PCA performed on this sample. Due to the lack of statistics, the only physically motivated PC is the first one. The overall shape of LRGs spectra in the observed frame is clearly visible in the first PC. The agreement between observed and simulated LRGs samples is good as shown by the PC and by the coefficient distributions in figure \ref{fig:tortorelli_fig14} and \ref{fig:tortorelli_fig15}.

\section{Conclusions}
\label{section:conclusions}

The Ultra Fast Image Generator (UFig) is an image simulator that relies upon simple models of galaxy properties, to produce realistic astronomical images. The models are based on underlying distributions that need to be calibrated using existing data. In this work, we test the model described in~\cite{herbel17} by using the extensive photometry provided by the PAUS survey.

To be able to assess how well the simulator performs and to test the model, it is important to perform the same data reduction steps on real and simulated images. Therefore, we develop a fast data analysis pipeline that can be consistently applied to data and simulations. The data analysis is performed on the set of PAUS images belonging to the COSMOS field, for a total of 2400 NB images. To extract source parameters, we perform forced photometry with \textsc{SE}, using Subaru images to detect sources and then create the detection images for PAUS from the detection broad-band catalogues.

To check the robustness of our methodology and to assess the performance of the image simulator, different diagnostics, both at the image and at the catalogue level, have been explored. For images and single-band catalogues, we find a good agreement in the distribution of pixel values, of magnitudes and in the magnitude-size relation. Furthermore, we check also inter-band correlations and find good agreement in the magnitude-magnitude and colour-magnitude distributions. We also check whether spectra extracted from simulations are realistic, comparing them with real ones.

The usual single and inter-band correlations are not well suited for a high dimensional dataset such as the PAUS one, therefore to capture the global information, we perform a Principal Component analysis (PCA). We derive the first 5 principal components (PCs) and we compare them. The first three PCs, which have the highest variance, agree very well, while the other two components show only a moderate agreement. The first two components capture the population of red galaxies, while the third component those galaxies characterized by recent star-formation. 

We then quantify the agreement between the PCs by creating the `mixing' matrix and comparing the coefficient distributions of the scalar product between the PAUS real and simulated PCs and both real and simulated data. We find that the mixing matrix numerically confirms the agreement between the first three PCs. Furthermore, the coefficient distributions show good agreement with a slight discrepancy in the contours involving the first coefficient. This will be improved in a future work by fitting the model using the approximate Bayesian computation (ABC) method on the full PAUS dataset.

Finally, we also select a sample of LRGs for both data and simulations. We perform PCA on this sample, finding that the overall shape of LRGs spectra in the observed frame is captured by the first PC.

We find that, even without further tuning, the model in~\cite{herbel17}, derived using broad-band data, gives very good results applied directly to this new higher-dimensional PAUS dataset. Future work will be devoted to the improvement of the model underlying our image simulator and the exploitation of galaxy population statistical properties by using the ABC method on the full PAUS dataset.

\acknowledgments

We acknowledge support by Swiss National Science Foundation (SNF) grant 200021\_169130. Funding for PAUS has been provided by Durham University (via the ERC StG DEGAS-259586), ETH Zurich, Leiden University (via ERC StG ADULT-279396 and Netherlands Organisation for Scientific Research (NWO) Vici grant 639.043.512) and University College London. The PAUS participants from Spanish institutions are partially supported by MINECO under grants CSD2007-00060, AYA2015-71825, ESP2015-66861, FPA2015-68048, SEV-2016-0588, SEV-2016-0597, and MDM-2015-0509, some of which include ERDF funds from the European Union. IEEC and IFAE are partially funded by the CERCA program of the Generalitat de Catalunya. The PAU data center is hosted by the Port d'Informaci\'o Cient\'ifica (PIC), maintained through a collaboration of CIEMAT and IFAE, with additional support from Universitat Aut\`onoma de Barcelona and ERDF. This work has made use of data from the European Space Agency (ESA) mission {\it Gaia} (\url{https://www.cosmos.esa.int/gaia}), processed by the {\it Gaia} Data Processing and Analysis Consortium (DPAC, \url{https://www.cosmos.esa.int/web/gaia/dpac/consortium}). Funding for the DPAC has been provided by national institutions, in particular the institutions participating in the {\it Gaia} Multilateral Agreement.


\bibliographystyle{unsrt}
\bibliography{tortorelli_jcap_bibliography}

% Please avoid comments such as "For a review'', "For some examples",
% "and references therein" or move them in the text. In general,
% please leave only references in the bibliography and move all
% accessory text in footnotes.

% Also, please have only one work for each \bibitem.

\appendix

\section{Coordinates of PAUS Fields}
\label{appendix:pauscoordinates}

We report in Table ~\ref{table:tortorelli_table1} the coordinates of the PAUS fields used in our analysis. The area covered corresponds to the COSMOS field.

\begin{table}
\centering
\begin{tabular}{c|c}
\hline
\textbf{RA (J2000)} & \textbf{DEC (J2000)} \\
\hline
09h 57m & 01$^{\circ}$ 33$^{'}$\\
09h 57m & 01$^{\circ}$ 41$^{'}$\\
09h 57m & 01$^{\circ}$ 49$^{'}$\\
09h 57m & 01$^{\circ}$ 57$^{'}$\\
09h 57m & 02$^{\circ}$ 05$^{'}$\\
09h 57m & 02$^{\circ}$ 13$^{'}$\\
09h 57m & 02$^{\circ}$ 21$^{'}$\\
09h 57m & 02$^{\circ}$ 29$^{'}$\\
09h 57m & 02$^{\circ}$ 37$^{'}$\\
09h 57m & 02$^{\circ}$ 45$^{'}$\\
09h 57m & 02$^{\circ}$ 53$^{'}$\\
09h 57m & 03$^{\circ}$ 01$^{'}$\\
09h 58m & 01$^{\circ}$ 33$^{'}$\\
09h 58m & 01$^{\circ}$ 41$^{'}$\\
09h 58m & 01$^{\circ}$ 49$^{'}$\\
09h 58m & 01$^{\circ}$ 57$^{'}$\\
09h 58m & 02$^{\circ}$ 05$^{'}$\\
09h 58m & 02$^{\circ}$ 13$^{'}$\\
09h 58m & 02$^{\circ}$ 21$^{'}$\\
09h 58m & 02$^{\circ}$ 29$^{'}$\\
09h 58m & 02$^{\circ}$ 37$^{'}$\\
09h 58m & 02$^{\circ}$ 45$^{'}$\\
09h 58m & 02$^{\circ}$ 53$^{'}$\\
09h 58m & 03$^{\circ}$ 01$^{'}$\\
09h 59m & 01$^{\circ}$ 33$^{'}$\\
09h 59m & 01$^{\circ}$ 41$^{'}$\\
09h 59m & 01$^{\circ}$ 49$^{'}$\\
09h 59m & 01$^{\circ}$ 57$^{'}$\\
09h 59m & 02$^{\circ}$ 05$^{'}$\\
09h 59m & 02$^{\circ}$ 13$^{'}$\\
09h 59m & 02$^{\circ}$ 21$^{'}$\\
09h 59m & 02$^{\circ}$ 29$^{'}$\\
09h 59m & 02$^{\circ}$ 37$^{'}$\\
09h 59m & 02$^{\circ}$ 45$^{'}$\\
09h 59m & 02$^{\circ}$ 53$^{'}$\\
09h 59m & 03$^{\circ}$ 01$^{'}$\\
10h 00m & 01$^{\circ}$ 33$^{'}$\\
10h 00m & 01$^{\circ}$ 41$^{'}$\\
10h 00m & 01$^{\circ}$ 49$^{'}$\\
10h 00m & 01$^{\circ}$ 57$^{'}$\\
10h 00m & 02$^{\circ}$ 05$^{'}$\\
10h 00m & 02$^{\circ}$ 13$^{'}$\\
10h 00m & 02$^{\circ}$ 21$^{'}$\\
10h 00m & 02$^{\circ}$ 29$^{'}$\\
10h 00m & 02$^{\circ}$ 37$^{'}$\\
10h 00m & 02$^{\circ}$ 45$^{'}$\\
\hline
\end{tabular}
\caption{The table shows the list of the PAUS fields used in our work. The fields are inside the COSMOS one.}
\label{table:tortorelli_table1}
\end{table}

\begin{table}
\centering
\begin{tabular}{c|c}
\hline
\textbf{RA (J2000)} & \textbf{DEC (J2000)} \\
\hline
10h 00m & 02$^{\circ}$ 53$^{'}$\\
10h 00m & 03$^{\circ}$ 01$^{'}$\\
10h 01m & 01$^{\circ}$ 33$^{'}$\\
10h 01m & 01$^{\circ}$ 41$^{'}$\\
10h 01m & 01$^{\circ}$ 49$^{'}$\\
10h 01m & 01$^{\circ}$ 57$^{'}$\\
10h 01m & 02$^{\circ}$ 05$^{'}$\\
10h 01m & 02$^{\circ}$ 13$^{'}$\\
10h 01m & 02$^{\circ}$ 21$^{'}$\\
10h 01m & 02$^{\circ}$ 29$^{'}$\\
10h 01m & 02$^{\circ}$ 37$^{'}$\\
10h 01m & 02$^{\circ}$ 45$^{'}$\\
10h 01m & 02$^{\circ}$ 53$^{'}$\\
10h 01m & 03$^{\circ}$ 01$^{'}$\\
10h 02m & 01$^{\circ}$ 33$^{'}$\\
10h 02m & 01$^{\circ}$ 41$^{'}$\\
10h 02m & 01$^{\circ}$ 49$^{'}$\\
10h 02m & 01$^{\circ}$ 57$^{'}$\\
10h 02m & 02$^{\circ}$ 05$^{'}$\\
10h 02m & 02$^{\circ}$ 13$^{'}$\\
10h 02m & 02$^{\circ}$ 21$^{'}$\\
10h 02m & 02$^{\circ}$ 29$^{'}$\\
10h 02m & 02$^{\circ}$ 37$^{'}$\\
10h 02m & 02$^{\circ}$ 45$^{'}$\\
10h 02m & 02$^{\circ}$ 53$^{'}$\\
10h 02m & 03$^{\circ}$ 01$^{'}$\\
10h 03m & 01$^{\circ}$ 33$^{'}$\\
10h 03m & 01$^{\circ}$ 41$^{'}$\\
10h 03m & 01$^{\circ}$ 49$^{'}$\\
10h 03m & 01$^{\circ}$ 57$^{'}$\\
10h 03m & 02$^{\circ}$ 05$^{'}$\\
10h 03m & 02$^{\circ}$ 13$^{'}$\\
10h 03m & 02$^{\circ}$ 21$^{'}$\\
10h 03m & 02$^{\circ}$ 29$^{'}$\\
10h 03m & 02$^{\circ}$ 37$^{'}$\\
10h 03m & 02$^{\circ}$ 45$^{'}$\\
10h 03m & 02$^{\circ}$ 53$^{'}$\\
10h 03m & 03$^{\circ}$ 01$^{'}$\\
\hline
\end{tabular}
\caption{Continue.}
\label{table:tortorelli_table2}
\end{table}

\section{Functioning Point for UFig Model}
\label{appendix:functioningpoint}

We report in table~\ref{table:tortorelli_table3} the specific set of model parameters (functioning point) we use in our paper. The Luminosity function (LF) is parametrized according to a Schechter function
\begin{equation}
\Phi (\mathrm{M,z}) = \frac{2}{5} \log{(10)} \phi_* 10^{\frac{2}{5} (\mathrm{M_* - M}) (\alpha + 1)} \exp{\left[ -10^{\frac{2}{5} (\mathrm{M_* - M})} \right]} 
\end{equation}
where the evolution with redshift of M$_*$ and $\phi_*$ is parametrized as
\begin{equation}
\begin{split}
\mathrm{M_*(z)} &= \mathrm{M_{*,slope}\ z + M_{*,intcpt} } \\
\mathrm{\phi_*(z)} &= \mathrm{\phi_{*,exp} \exp{(\phi_{*,amp}\ z)}} \\
\end{split}
\end{equation}
 The parameters for the populations of blue and red galaxies are taken from \cite{beare15}. The model of galaxy sizes is characterized by a log-normal distribution of the physical radius r$_{50}^{\mathrm{phys}}$ with fixed standard deviation $\mathrm{r_{50,std}^{phys}}$ for all galaxies and with mean $\mu_{\mathrm{phys}}$, which depends linearly on the absolute magnitude M drawn from the LF
\begin{equation}
\mu_{\mathrm{phys}} (M) = \mathrm{r_{50,slope}^{phys}}\ M + \mathrm{r_{50,intcpt}^{phys}}
\end{equation}
The set of sizes used to determine the best-fitting coefficients is from the Great-3 challenge \cite{great-3}. The coefficients of the basis spectra, instead, are empirically motivated in \cite{herbel17}.

\begin{table}
\centering
\begin{tabular}{c c c}
\hline
& \textbf{Red} & \textbf{Blue} \\
\hline
$\alpha$ & -0.5 & -1.3 \\
\hline
$\mathrm{M_{*,slope}}$ & -0.70798041 & -0.9408582 \\
\hline
$\mathrm{M_{*,intcpt}}$ & -20.37196157 & -20.40492365 \\
\hline
$\mathrm{\phi_{*,exp}}$ & -0.70596888 & -0.10268436 \\
\hline
$\mathrm{\phi_{*,amp}}$ & 0.0035097 & 0.00370253 \\
\hline
$\mathrm{r_{50,slope}^{phys}}$ & -0.24293465 & -0.24293465 \\
\hline
$\mathrm{r_{50,intcpt}^{phys}}$ & 1.2268735 & 1.2268735 \\
\hline
$\mathrm{r_{50,std}^{phys}}$ & 0.56800081 & 0.56800081 \\
\hline
$\mathrm{\alpha_1}$ & 1.62158197 & 1.9946549 \\
\hline
$\mathrm{\alpha_2}$ & 1.62137391 & 1.99469164\\
\hline
$\mathrm{\alpha_3}$ & 1.62175061 & 1.99461187\\
\hline
$\mathrm{\alpha_4}$ & 1.62159144 & 1.9946589\\
\hline
$\mathrm{\alpha_5}$ & 1.62165971 & 1.99463069\\
\hline
\end{tabular}
\caption{Functioning Point used in this work for red and blue galaxies.}
\label{table:tortorelli_table3}
\end{table}

\section{\textsc{Source Extractor} Configuration}
\label{appendix:sexconfig}

We report in table~\ref{table:tortorelli_table4} the \textsc{Source Extractor} configuration that we use to analyze Subaru and PAUS observed and simulated images. The same configuration is used for \textsc{Source Extractor} single image and dual image mode.

\begin{table}
\centering
\begin{tabular}{c|c}
\hline
\textbf{\textsc{Source Extractor} parameter name} & \textbf{Value} \\
\hline
CATALOG\_TYPE & FITS\_1.0\\
DETECT\_TYPE & CCD\\
DETECT\_MINAREA & 5\\
DETECT\_THRESH & 1.7\\
ANALYSIS\_THRESH & 1.7\\
FILTER & Y\\
FILTER\_NAME & gauss\_3.0\_5x5.conv\\
DEBLEND\_NTHRESH & 64\\
DEBLEND\_MINCONT & 0.0001\\
CLEAN & Y\\
CLEAN\_PARAM & 1.0\\
MASK\_TYPE & CORRECT\\
PHOT\_APERTURES & 5, 10, 15, 20, 25\\
PHOT\_AUTOPARAMS & 2.5, 3.5 \\
PHOT\_FLUXFRAC & 0.5\\
SATUR\_LEVEL & tile- \& band-dependent (Subaru), 262143 (PAUS)\\
MAG\_ZEROPOINT & 31.4 (Subaru), tile- \& band-dependent (PAUS)\\
GAIN & band-dependent\\
PIXEL\_SCALE & 0.150 (Subaru), 0.265 (PAUS)\\
SEEING\_FWHM & tile- \& band-dependent\\
STARNNW\_NAME & default.nnw\\
BACK\_SIZE & 32\\
BACK\_FILTERSIZE & 3\\
BACKPHOTO\_TYPE & LOCAL\\
BACKPHOTO\_THICK & 24\\
WEIGHT\_TYPE & MAP\_RMS (Subaru), NONE (PAUS)\\
CHECKIMAGE\_TYPE & SEGMENTATION, APERTURES\\
\hline
\end{tabular}
\caption{\textsc{Source Extractor} configuration used in this work.}
\label{table:tortorelli_table4}
\end{table}

\end{document}